\documentclass[twocolumn,showpacs,preprintnumbers,amsmath,amssymb,aps]{revtex4}
\usepackage{graphicx}
\begin{document}

\title{Vortex Molecular Crystal and Vortex Plastic Crystal
States in Honeycomb and Kagom{\' e} 
Pinning Arrays} 
\author{C. Reichhardt  
and C.J. Olson Reichhardt} 
\affiliation{ 
Theoretical Division and Center for Nonlinear Studies,
Los Alamos National Laboratory, Los Alamos, New Mexico 87545}
\date{\today}
\begin{abstract}

Using numerical simulations, 
we investigate vortex configurations and pinning in 
superconductors with honeycomb and kagom{\' e} pinning arrays.
We find that a variety of novel vortex crystal states can be stabilized at
integer and fractional matching field densities.  
The honeycomb and kagom{\' e} pinning arrays produce
considerably more pronounced commensuration peaks
in the critical depinning force than triangular pinning
arrays, and also cause additional peaks at 
noninteger matching fields where
a portion of the vortices are located in the large interstitial regions
of the pinning lattices. 
For the honeycomb pinning array,
we find matching effects of 
equal strength at most fillings $B/B_{\phi} = n/2$ for $n > 2$, where
$n$ is an integer,
in agreement with recent experiments. 
For kagom{\' e} pinning arrays, pronounced matching effects generally occur at 
$B/B_{\phi} = n/3$ for $n > 3$, while for triangular pinning arrays pronounced 
matching effects are observed only at integer fillings $B/B_{\phi}=n$.    
At the noninteger matching field peaks
in the honeycomb and kagom{\' e} pinning arrays,
the interstitial vortices
are arranged in 
dimer, trimer, and higher order $n$-mer states that have an overall 
orientational order. 
We call these $n$-mer states
``vortex molecular crystals'' and ``vortex plastic crystals'' 
since they are similar to the states 
recently observed in colloidal molecular crystal systems. 
We argue that the vortex molecular
crystals have properties in common with
certain spin systems such as Ising and n-state Potts models.     
We show that kagom{\' e} 
and honeycomb pinning arrays can be useful for
increasing the critical current above that of purely 
triangular pinning arrays. 
\end{abstract}
\maketitle

\vskip2pc

\section{Introduction}
There have been extensive   
studies on the static and dynamical properties of vortices 
in superconductors with periodic arrays of artificial pinning sites. 
These works focused on simple two-dimensional 
periodic pinning arrays such as 
square, triangular, and rectangular
lattices, where the pinning sites 
consist of holes 
\cite{Fiory,Metlushko,Baert,Harada,Field, Welp,Reichhardt,Jensen1,Peeters},
blind holes \cite{Pannetier,Raedts},
or magnetic dots 
\cite{Martin,Morgan,VanBael,Martin99,Bending,Fertig,Chen}. 
For square and triangular pinning arrays,
pronounced commensurability effects such as 
peaks or anomalies in the critical current appear at the magnetic field
$B=B_\phi$ where the number of vortices equals the number of pinning sites,
as well as at higher fields $B=nB_\phi$, where $n$ is an integer. 
At these matching fields 
the vortex lattice forms ordered crystalline structures of a type
determined by the number of vortices that
are captured at individual pinning sites. 
If more than one vortex can occupy each
pinning site in the form of a multi-quanta vortex,  
then at each matching field the 
overall vortex lattice has the same symmetry as the pinning lattice
but is composed of $n$-quanta vortices.
If only a single vortex can occupy each pinning site, 
ordered vortex crystals still form at
the matching fields and for $n>1$ some of the vortices are located in the
interstitial regions between the 
pinning sites. Imaging experiments \cite{Harada} and
simulations \cite{Reichhardt}
for systems where at most one vortex can occupy each pin have shown
that numerous kinds of interstitial vortex lattice structures can
be stabilized, some of which have different symmetries than the pinning
array.
Interstitial vortex crystals also form above the pinning saturation field
in samples with pins that can be occupied by multi-quanta vortices.
When each pin has captured as many vortices as possible, additional
vortices sit in the interstitial regions, and the resulting 
vortex lattice
structure is a composite of interstitial singly-quantized and 
pinned multiply-quantized vortices
\cite{Metlushko,Peeters,Karapetrov,Olson}. 
In the case where the pinning sites are blind holes, it is possible
for multiple vortices to occupy a single pinning site without merging
into a multi-quanta vortex.  Instead, the vortices retain their
individual identities and form dimer, trimer, or ring-type states
inside the pin \cite{Pannetier, Jensen}.

In addition to the
matching effects that appear at integer fields, 
commensuration effects can also occur
at nonmatching fields or fractional fields \cite{Baert,Field,Jensen1,Bending}.
These noninteger matching effects are generally 
weaker than those observed at integer matching and are most prominent 
for fields $B<B_\phi$.
If multiple vortex quantization occurs above the first matching field, the
sub-matching sequence that appears between $B=0$ and $B=B_\phi$ is
repeated between every integer matching field until the pinning sites
are saturated.
In contrast, 
if only one vortex is captured per pinning site the fractional matching
effects above the first matching field are 
significantly reduced or missing \cite{Jensen1}.
Commensuration effects for vortices interacting with a periodic substrate 
have also recently been demonstrated 
for vortices in  
Bose-Einstein condensates where the 
pinning sites are created with an optical array
\cite{Bigelow,Demler,Tung}. 

The physics of vortices in periodic pinning arrays is similar to that
of repulsively interacting colloids in triangular or square 
periodic trap arrays  
\cite{Colloid,Brunner,Trizac,Frey,Mangold} 
and charged spheres on periodic substrates \cite{Guthermann}. 
In both these cases it is possible to have localized traps which 
capture only a single colloid or sphere while the remaining 
particles sit in the interstitial regions \cite{Mangold,Guthermann},
similar to the situation for vortex pinning arrays.
The interstitial particles are more mobile than the pinned particles, 
and the particle trajectories resemble those seen in 
computer simulations of vortices in similar geometries 
\cite{Dominguez,Zimanyi}.
In the colloid system it is also possible for multiple colloids to
be captured by a single trap
\cite{Colloid,Brunner,Trizac,Frey} where they form dimer, trimer, or
higher order $n$-mer states. 
The $n$-mers may be orientationally ordered if the interaction between
colloids in neighboring traps is strong enough, and the system
can show multistage melting transitions in which the initial stage 
of melting occurs when the orientational ordering of the $n$-mers is lost,
followed by a transition in which the $n$-mers break apart.
The ordered $n$-mer states can be mapped onto various
spin systems such as Ising and Potts models \cite{Trizac,Frey}. 
Orientationally 
ordered dimer and trimer states have also been proposed 
to occur for vortices in superconductors with blind hole arrays 
where the vortices retain single quantization in an individual hole
\cite{Jensen1}. 
The orientational 
ordering of the $n$-mer states in the colloidal and 
vortex systems arises due to quadrupole or higher order pole moment
interactions between the $n$-mers in adjacent traps \cite{Trizac}.
The anisotropic nature of the quadrupole and higher order pole moments
creates preferred directions for the alignment of the $n$-mers that
minimize the pole moment energy.

\begin{figure}
\includegraphics[width=3.5in]{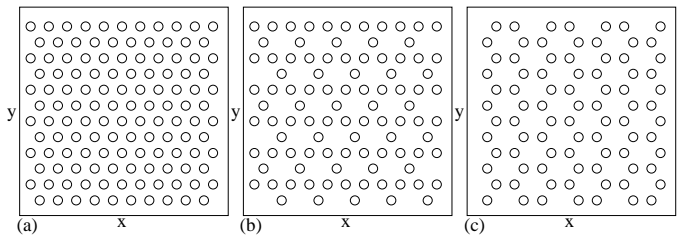}
\caption{
The pinning site locations (circles) for periodic arrangements of sites
forming (a) triangular,  
(b) kagom{\' e}, and (c) honeycomb arrays. 
Here $B_\phi = 0.47\phi_0/\lambda^2$.  
}
\label{fig:pinimage}
\end{figure}

A 
limited number of studies have 
treated
periodic substrates other than square and triangular pinning arrays,
such as honeycomb or kagom{\' e} pinning arrays. 
These are simply
triangular lattices with a fraction of the pinning sites removed. 
Figure \ref{fig:pinimage}(a) shows a triangular pinning lattice.
Removing every other pinning site from every other row of the triangular
lattice, which eliminates 1/4 of the pinning sites, 
produces the kagom{\' e} array illustrated in 
Fig.~\ref{fig:pinimage}(b). 
A honeycomb array in which 1/3 of the pinning sites in the triangular
array have been eliminated is shown in Fig.~\ref{fig:pinimage}(c). 
Here every third pinning site is removed from every row 
of the triangular array.
An experimental study of pinning phenomena in kagom{\' e} arrays 
using superconductors with magnetic dot arrays 
produced evidence for pronounced commensurability effects 
at noninteger matching fields \cite{Morgan}.  These results are, however,
difficult to interpret since magnetic dot arrays 
can induce the formation of vortices or antivortices
in addition to the vortices created by the
external field \cite{Fertig}.   
Numerical work on kagom{\' e} pinning arrays \cite{Dominguez}
only treated thermal melting of vortices at a
single field of $B/B_{\phi} = 2.0$ where 
a two step melting transition was shown to occur.
Here, the multiple interstitial vortices located at the 
larger interstitial sites where a pin has been removed undergo 
local melting at a temperature below that at which
the entire vortex lattice melts.
In Ref.~\cite{Dominguez}, neither long range vortex configurations nor
critical currents were analyzed, so the nature of
commensurability effects for varied vortex densities in kagom{\' e} pinning
arrays is not known.    

Only a single experimental study \cite{Wu}
has been performed on honeycomb pinning
arrays to our knowledge.
In Ref.~\cite{Wu}, several unusual features were observed, including
pronounced matching effects of equal magnitude 
at magnetic fields corresponding to both integer and half integer 
multiples of the matching field for vortex densities up to
to the fifth matching field $B/B_\phi=5$.   
This result is in contrast to 
the response of square or triangular pinning arrays, where 
commensuration effects are much weaker 
at nonmatching fields than at integer matching fields.
In Ref.~\cite{Wu}, it was also observed that for
fields greater than the second matching field, $B/B_\phi>2$, the 
commensuration effects at half-integer matching fields
become more prominent than those at integer matching fields.
This effect can be understood by considering that since the honeycomb
pinning array is simply a triangular pinning 
array that has been diluted by 1/3, 
the field $B/B_\phi=1.5$ for the honeycomb array would correspond to an
integer matching field $B/B_\phi=1$ in a triangular array of the same density.
As a result, the overall vortex lattice is triangular at $B/B_\phi=1.5$ in
a honeycomb pinning array.
Similarly, a field  of $B/B_\phi=1/2$ in the honeycomb pinning array 
would correspond to a field of $B/B_\phi=1/3$ 
in the equivalent triangular pinning array, 
which is known to produce a peak in the critical current \cite{Jensen1}. 
These results suggest that honeycomb pinning arrays may 
allow for a variety of new vortex structures
to be stabilized at noninteger fillings. 

In this work we present the first extensive study of 
vortex pinning and dynamics in honeycomb and kagom{\' e}
pinning arrays using numerical simulations.  Section II contains
a description of the simulation method.  
We show the vortex configurations and ordering for honeycomb
pinning arrays in Section III and illustrate the formation of vortex
molecular crystals, which are named in analogy with molecular crystals.
Section IV gives the corresponding description for 
vortices on a kagom{\' e} pinning lattice.
The melting of these vortex configurations and the creation of vortex plastic
crystal states for both types of pinning lattices is studied in Section V.
We construct phase diagrams for the vortex molecular crystals at
the dimer and trimer fillings of the honeycomb pinning lattice 
as a function of temperature and pinning strength in Section VI.
The effect of the strength of the pinning sites is explored in further
detail in Section VII.  The paper closes with a discussion in Section VIII and a
conclusion in Section IX.

\section{Simulation}
We perform two-dimensional simulations of superconducting vortices
in honeycomb and kagom{\' e} pinning arrays
using a computational procedure 
similar to that previously employed for studies of 
vortices in square and triangular pinning
arrays \cite{Reichhardt,Jensen1,Jensen,Olson}.  
The system of size $L_x \times L_y$ has periodic boundary conditions in 
the $x$ and $y$ directions and 
contains $N_p$ 
pinning sites and $N_{v}$ vortices. 
The  dynamics of a vortex $i$ located at position ${\bf R}_i$ 
is determined by the 
following overdamped equation of motion: 
\begin{equation}
\eta\frac{ d{\bf R}_{i}}{dt} = {\bf F}_{i}^{vv} + {\bf F}^{p}_{i}  
+ {\bf F}^{d}_{i} + {\bf F}^{T}_{i} .
\end{equation}
Here ${\bf F}_i^{vv}$ is the repulsive vortex-vortex interaction force,  
${\bf F}^p_i$ is the force from the pinning sites,
${\bf F}^d_i$ is the force from an external drive,
and ${\bf F}^T_i$ is the random force from thermal fluctuations.
The damping constant $\eta=\phi_0^2d/2\pi\xi^2\rho_N$ where 
$d$ is the thickness of the superconducting sample, 
$\phi_0=h/2e$ is the flux quantum, $\xi$ is the superconducting
coherence length, and $\rho_N$ is the normal state resistivity of the
material \cite{Tinkham}.  We measure length in units of the London penetration
depth $\lambda$
and for most of the results presented here the system size is
$24\lambda\times 24\lambda$. 

The explicit vortex-vortex interaction force is 
\begin{equation}
{\bf F}_{i}^{vv} = \sum^{N_{v}}_{j\ne i}f_{0}
K_{1}\left(\frac{R_{ij}}{\lambda}\right){\bf {\hat R}}_{ij} , 
\end{equation}
where $K_{1}$ is the modified Bessel function, 
$f_{0}=\phi_0^2/(2\pi\mu_0\lambda^3)$,
$R_{ij}=|{\bf R}_i-{\bf R}_j|$, and
${\bf {\hat R}}_{ij}=({\bf R}_i-{\bf R}_j)/R_{ij}$.
A short range cutoff of $0.1\lambda$ is
applied to the vortex-vortex interaction force to avoid divergence; 
however, at
the densities considered here, vortices do not approach each other this
closely.
The interaction is also cut off beyond $6\lambda$ for computational
efficiency since the vortex-vortex forces beyond this 
distance are negligible \cite{Reichhardt}. 
Time is measured in units of $\tau=\eta/f_0$.
As an example of physical units, for
a NbSe$_2$ crystal of thickness $d=0.1$ mm
with $\eta=2.36\times 10^{-11}$ Ns/m, 
$f_{0}=6.78\times 10^{-5}$ N/m and
$\tau=0.35$ $\mu$s.

The pinning sites are modeled as $N_p$ attractive parabolic traps 
with radius $R_{p}=0.3\lambda$ and strength $f_{p}$, so that
\begin{equation}
{\bf F}^{p}_{i} =  -\sum_{k=1}^{N_p}\frac{f_{p}}{R_{p}}R^{(p)}_{ik}
\Theta\left(\frac{R_{p} - R^{(p)}_{ik}}{\lambda}\right)
{\bf {\hat R}}^{(p)}_{ik}
\end{equation}
Here ${\bf R}^{(p)}_{k}$ is the location of pinning site $k$, 
$R^{(p)}_{ik}=|{\bf R}_{i} - {\bf R}^{(p)}_{k}|$,
${\bf {\hat R}}^{(p)}_{ik}=({\bf R}_{i} - {\bf R}^{(p)}_{k})/R^{(p)}_{ik}$,
and $\Theta$ is the Heaviside step function.
The pinning radius is chosen to be small enough that only one vortex can
be captured per pinning site.
The pinning sites are arranged in a triangular lattice with lattice
constant $a_0$, as illustrated in Fig.~\ref{fig:pinimage}(a), and
then 1/4 or 1/3 of the pinning sites are removed to create a kagom{\' e}
or honeycomb array, as in Fig.~\ref{fig:pinimage}(b,c).
The pinning density is $n_p=N_p/(L_xL_y)$. 
The matching fields for the honeycomb and kagom{\' e} arrays are
$B_\phi^H=n_p$ and $B_\phi^K=n_p$, respectively, while the matching
field of the equivalent triangular array 
is $B_\phi$, such that $B_\phi^H/B_\phi=2/3$ and
$B_\phi^K/B_\phi=3/4$.

The driving force is assumed to arise
from the application of an external current which induces a  
Lorentz force on the vortices that is perpendicular to the current. 
All vortices experience an equal driving force ${\bf F}^{d}=F^d{\bf {\hat x}}$ 
in the $x$ direction, 
corresponding to the horizontal axis of Fig.~\ref{fig:pinimage}. 
The thermal force ${\bf F}^{T}_{i}$
is modeled as random Langevin kicks with the properties
$\langle{\bf F}^{T}_{i}\rangle = 0$ and
$\langle{\bf F}_{i}^{T}(t){\bf F}_{j}^{T}(t^{\prime})\rangle 
= 2\eta k_{B}T \delta(t-t^{\prime})\delta_{ij}$. 

The initial vortex configurations are obtained by simulated annealing. 
Our procedure for this study was to start from an initial temperature of
$F^T=3$ and decrease the temperature 
to $F^T=0$ in increments of $\delta F^T=0.002$ while
spending 5000 simulation time steps at each increment, so that the total
annealing time is $7.5 \times 10^6$ simulation time steps.
We note that an overly rapid annealing rate can cause the system to be
trapped in a metastable state, which prevents the vortices from ordering
even at integer matching fields.
To check our annealing rate, we tested slower rates and found that the
resulting vortex configurations were unchanged.
Once the vortex positions have been initialized
and the temperature has been set to zero, we determine
the velocity-force curve relations
and the critical depinning force 
$f_c$ by slowly increasing the external drive $F^d$.
We measure the average vortex velocity 
response 
$V_x=N_v^{-1}\langle\sum_{i=1}^{N_v}{\bf v}_{i}\cdot {\bf {\hat x}}\rangle$,
where ${\bf v}_i=d{\bf R}_i/dt$.
The resulting velocity-force curve would correspond to a voltage-current curve
in experiment. The depinning force is determined
by applying a cut-off threshold of $V_x=0.01$
to the average velocity. We find that this cutoff is sufficiently low that
the critical depinning force $f_{c}$ vs $B$ curves 
are not strongly sensitive to the choice of cutoff.

\section{Vortex Pinning and Ordering in Honeycomb Arrays}

\subsection{Commensurability Peaks at Integer and Half Integer Fillings} 

\begin{figure}
\includegraphics[width=3.5in]{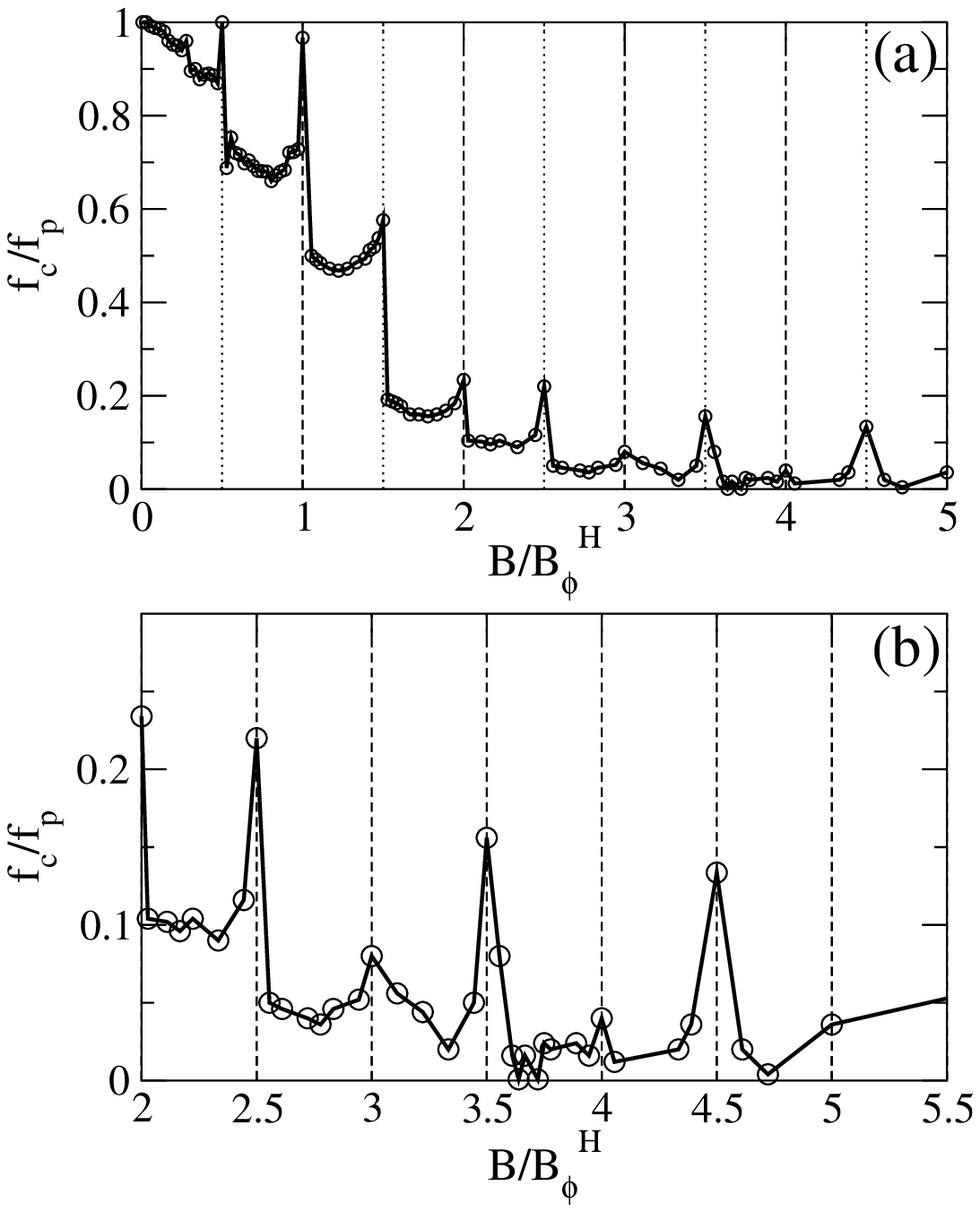}
\caption{
(a) The depinning force $f_{c}/f_{p}$ vs $B/B^H_{\phi}$ for the 
honeycomb pinning array illustrated in Fig.~\ref{fig:pinimage}(c)
with $f_p=0.5f_0$, $B_\phi=0.47\phi_0/\lambda^2$, 
and $B^H_\phi=0.313\phi_0/\lambda^2$. 
(b) A blow-up from (a) of $f_{c}/f_{p}$ vs $B/B^H_{\phi}$ 
in the region $B/B^H_\phi \ge 2$.
}
\label{fig:depinhoney}
\end{figure}

We measure the critical depinning force $f_c$ for vortices on a honeycomb
pinning array with $f_p=0.5f_0$ and $B_\phi=0.47\phi_0/\lambda^2$
as a function of vortex density. 
The results are plotted in Fig.~\ref{fig:depinhoney}(a)
as $f_c/f_p$ versus $B/B^H_\phi$
where $B^H_{\phi}=0.313\phi_0/\lambda^2$ is the field at which 
the number of vortices equals the number 
of pinning sites for the honeycomb pinning array.
We find peaks in $f_{c}$ at the integer matching fields
$B/B^H_{\phi} = 1$, 2, 3, and $4$.    
Additionally, there are pronounced peaks in $f_c$ at the half-matching fields
$B/B^H_{\phi} = 0.5$, 1.5, 2.5, 3.5, and $4.5$.
Note that the $f_c$ peaks at the noninteger matching 
fields of $B/B^H_{\phi}=3.5$ and $4.5$ are significantly larger than the 
peaks at the integer matching fields of $B/B^H_\phi=3$, $4$ and $5$, as
highlighted in Fig~\ref{fig:depinhoney}(b).
There are sub-matching commensuration effects for
$0.5 < B/B^H_\phi < 1.0$ which are in general weaker than the sub-matching 
commensuration effects for $B/B^H_\phi=0$ and 1.0. A similar trend was 
observed in earlier works for square and triangular pinning 
arrays \cite{Baert,Jensen1}. 
The numerical simulations are time consuming and permit us to perform
only finite field increments, so that the weaker submatching and higher
order fractional matching fields are difficult to observe numerically. 
In this paper we focus mainly on the pronounced matching fields.

\begin{figure}
\includegraphics[width=3.5in]{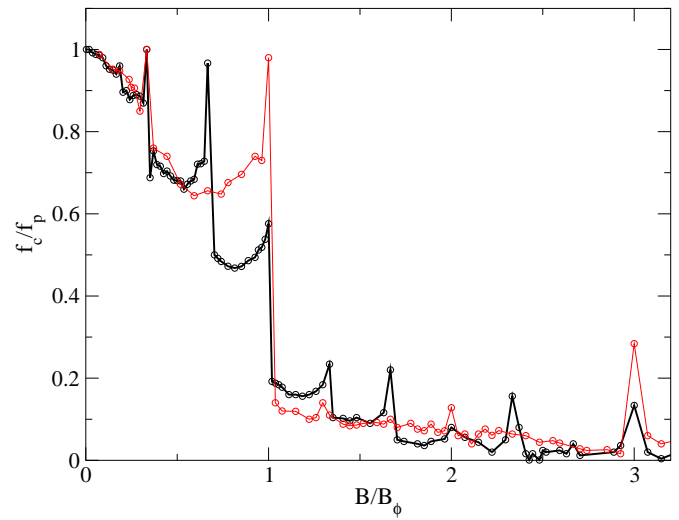}
\caption{(Color online)
$f_{c}/f_{p}$ vs $B/B_{\phi}$ for the 
honeycomb pinning array (black heavy line) and the triangular pinning array 
(red light line) in Fig.~1(a,c) with $f_p=0.5f_0$
and $B_\phi=0.47\phi_0/\lambda^2$.
}
\label{fig:honeytri}
\end{figure}

The behavior of the commensuration effects 
for the honeycomb pinning 
lattice in Fig.~\ref{fig:depinhoney} is very similar to the
experimental results observed for honeycomb arrays 
in Ref.~\cite{Wu}. In the experiments, 
strong commensuration effects appeared at 
$B/B^H_{\phi} = 1/2$, 1, 1.5, and $2$, in agreement
with our results. The experiments also show that
the commensuration effects at
$B/B^H_{\phi} = 3$ and $4$ are very weak or absent
while those at $B/B^H_{\phi} = 3.5$ and $4.5$ 
are very pronounced.  This is also in agreement 
with our result as seen in Fig.~\ref{fig:depinhoney}(b). 
One difference between the experimental results and our work 
is that we find a very pronounced peak in $f_c$ at
at $B/B^H_{\phi} = 2.5$, while the experiments produced only a weak peak
at the same field.
In Section V we show that this may be due to the 
fact that the commensurate vortex configuration at $B/B^H_{\phi} = 2.5$ 
is unstable under thermal fluctuations.
                  
Figure \ref{fig:honeytri} shows a comparison between 
the behavior of $f_c$ as a function of $B$ 
for the honeycomb pinning array (heavy line) and 
a triangular pinning array with the same value of 
$B_\phi=0.47\phi_0/\lambda^2$ (light line).
The depinning force $f_c$ for the triangular pinning array exhibits 
pronounced peaks at $B/B_{\phi}=1/3$, $2$, and $3$,
and only very weak peaks at 
noninteger matching fields for $B/B_{\phi} > 1$.
The peak at the second matching field $B/B_{\phi} = 2$ for the
triangular pinning array
is significantly 
smaller than the one at the third matching
field $B/B_\phi=3$. 
This effect has been observed in previous studies and results from 
the fact that the vortex lattice forms a honeycomb structure at
the second matching field while at the third matching field the overall
symmetry of the vortex lattice is triangular \cite{Reichhardt}.
Figure~\ref{fig:honeytri} 
also indicates that the peak at $B/B_{\phi} = 1.0$ for the triangular
array coincides with the peak at $B/B^H_{\phi} = 1.5$ for the honeycomb array, 
while a peak at $B/B_{\phi} = 1/3$ in the triangular array
coincides with a peak at $B/B^{H}_{\phi} = 1/2$ in the
honeycomb array.    

\subsection{Vortex Configurations for $B\le B_\phi$}  

\begin{figure}
\includegraphics[width=3.5in]{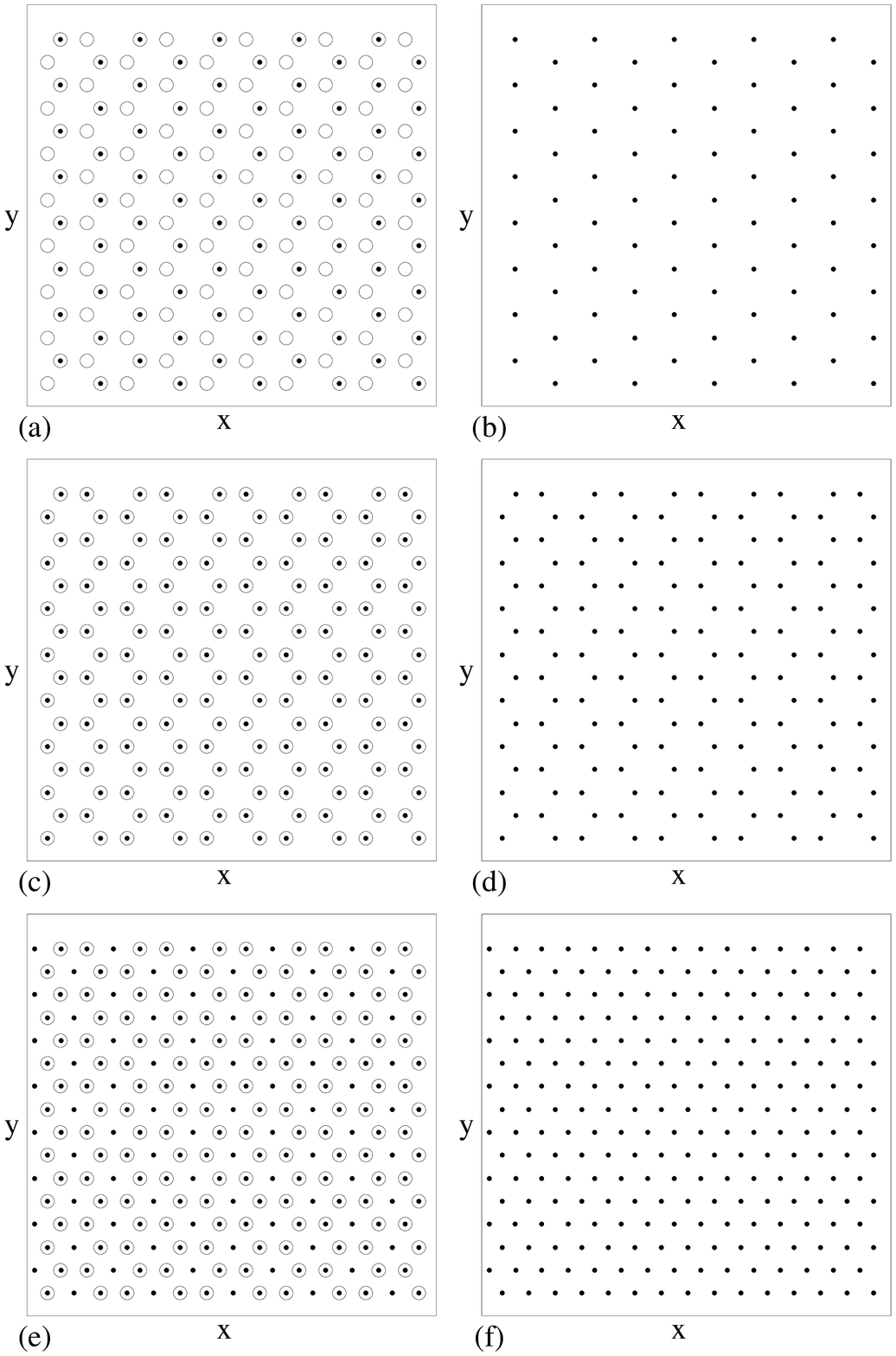}
\caption{
Left panels: 
Vortex positions (black dots) and pinning site positions (open circles) obtained 
for the honeycomb array in Fig.~\ref{fig:pinimage}(c) and 
Fig.~\ref{fig:depinhoney}. 
Right panels: Vortex positions only.
(a,b) $B/B^H_{\phi} = 1/2$. 
(c,d) $B/B^H_{\phi} = 1$. 
(e,f) $B/B^H_{\phi} =1.5$. 
}
\label{fig:honeyimg}
\end{figure}

In order to explain the origin of the pronounced commensurability peaks 
in $f_c$ for the honeycomb pinning array at $B/B^H_{\phi} = n/2$
and the fact that some peaks are more prominent than others, 
we analyze the vortex configurations at these fields.     
In the left panels of Fig.~\ref{fig:honeyimg} 
we illustrate the vortex and pinning site locations for three matching
fields in the honeycomb array from Fig.~\ref{fig:depinhoney}, while
the right panels of Fig.~\ref{fig:honeyimg} show only the vortex positions
at the same fields in order to emphasize the vortex lattice structure.

Figure \ref{fig:honeyimg}(a,b) indicates that
at $B/B^H_{\phi} = 1/2$ 
the overall vortex lattice is triangular. 
The vortex configuration at half filling for the honeycomb lattice
is the same as that of vortices at unit filling
in an undiluted triangular lattice with
matching field $B^H_\phi/2$ that is rotated by 90$^\circ$ relative
to the triangular lattice in Fig.~\ref{fig:pinimage}(a).
The configuration is also identical to the vortex arrangement
found for $1/3$ filling of a triangular lattice with matching
field $B_\phi$ \cite{Jensen1}.  
Since the vortex lattice structure
is triangular, the vortex-vortex interactions 
cancel and the depinning force is determined only
by the maximum force of the pinning sites such 
that $f_{c}/f_{p} = 1.0$. 
This is what we find at $B/B^H_\phi=1/2$, as seen in 
Fig.~\ref{fig:depinhoney}. 
For fields just above or below half filling, the vortex lattice
retains the same triangular ordering shown in Fig.~\ref{fig:honeyimg}(a,b) but 
contains weakly pinned vacancies or interstitials which reduce
the value of $f_c$.
In the case of a triangular pinning array, there 
is no commensurate peak at $B/B_{\phi} = 1/2$ when all of the vortices are
forced to occupy pinning sites since the system is geometrically frustrated,
resulting in a strongly defected vortex lattice \cite{Jensen1}.
If $f_{p}$ is very weak an ordered vortex lattice
can form at this field
when the elastic forces of the vortex lattice
overcome the pinning force, allowing half of the
vortices to shift out of the pinning sites and creating a partially pinned
or floating
triangular vortex lattice  \cite{Jensen1,Guthermann}. 
In our simulations the pinning strength $f_p=0.5f_0$ is well above this limit so 
that floating vortex configurations do not occur; we return to this point
in Sections VI and VII.
We note that for very low applied fields $B/B_\phi \ll 1$, the depinning
force is dominated by single vortex pinning and thus $f_c/f_p \approx 1.$

The vortex configuration for the first matching field in the honeycomb
array, $B/B^H_\phi=1$, is illustrated in Fig.~\ref{fig:honeyimg}(c,d).
The vortex lattice has the same highly symmetric structure as the
pinning lattice and as a result the vortex-vortex interactions 
cancel, giving $f_{c}/f_{p} = 1.0$ as 
shown in Fig.~\ref{fig:depinhoney}. 
For fields above the first matching field,
the additional $N_v-N_p$ vortices are located in the 
interstitial regions and are pinned not by the pinning sites
but by the interactions with the vortices trapped at the pinning sites. 
In general this interstitial pinning is weak,
so  $f_{c}/f_{p}$ drops significantly for $B/B^H_{\phi} > 1.0$ as
seen in Fig.~\ref{fig:depinhoney}. 

At $B/B^H_{\phi} = 1.5$,  where a  prominent peak in the depinning
force appears in Fig.~\ref{fig:depinhoney}, 
the overall vortex lattice is triangular as 
shown in Fig.~\ref{fig:honeyimg}(e,f). 
In this case the interstitial sites that were produced when the triangular
pinning array was diluted to form the honeycomb pinning array each
capture one vortex.
Figure~\ref{fig:honeytri}(b) shows that
the prominent peak in $f_{c}/f_{p}$ at $B/B_\phi=1$ 
for the triangular pinning array
coincides with the peak at $B/B^H_\phi=1.5$ in the
honeycomb pinning array.
Although the symmetry of the vortex lattice is triangular in each case,
at $B/B^H_\phi=1.5$ for the honeycomb pinning array the depinning force
is determined not by $f_{p}$ but by the
caging force on the vortices in the interstitial regions. 
As a result, the depinning force for the honeycomb array at 
this field is lower than the depinning force for the triangular array 
at $B/B_\phi=1.0$ filling, as seen in Fig.~\ref{fig:honeytri}(b). 
For fields slightly below or above
$B/B^H_{\phi} = 1.5$ in the honeycomb pinning array, 
interstitials or vacancies appear in the triangular vortex lattice at the
locations of the missing pins and cause a reduction in $f_c/f_p$.

\begin{figure}
\includegraphics[width=3.5in]{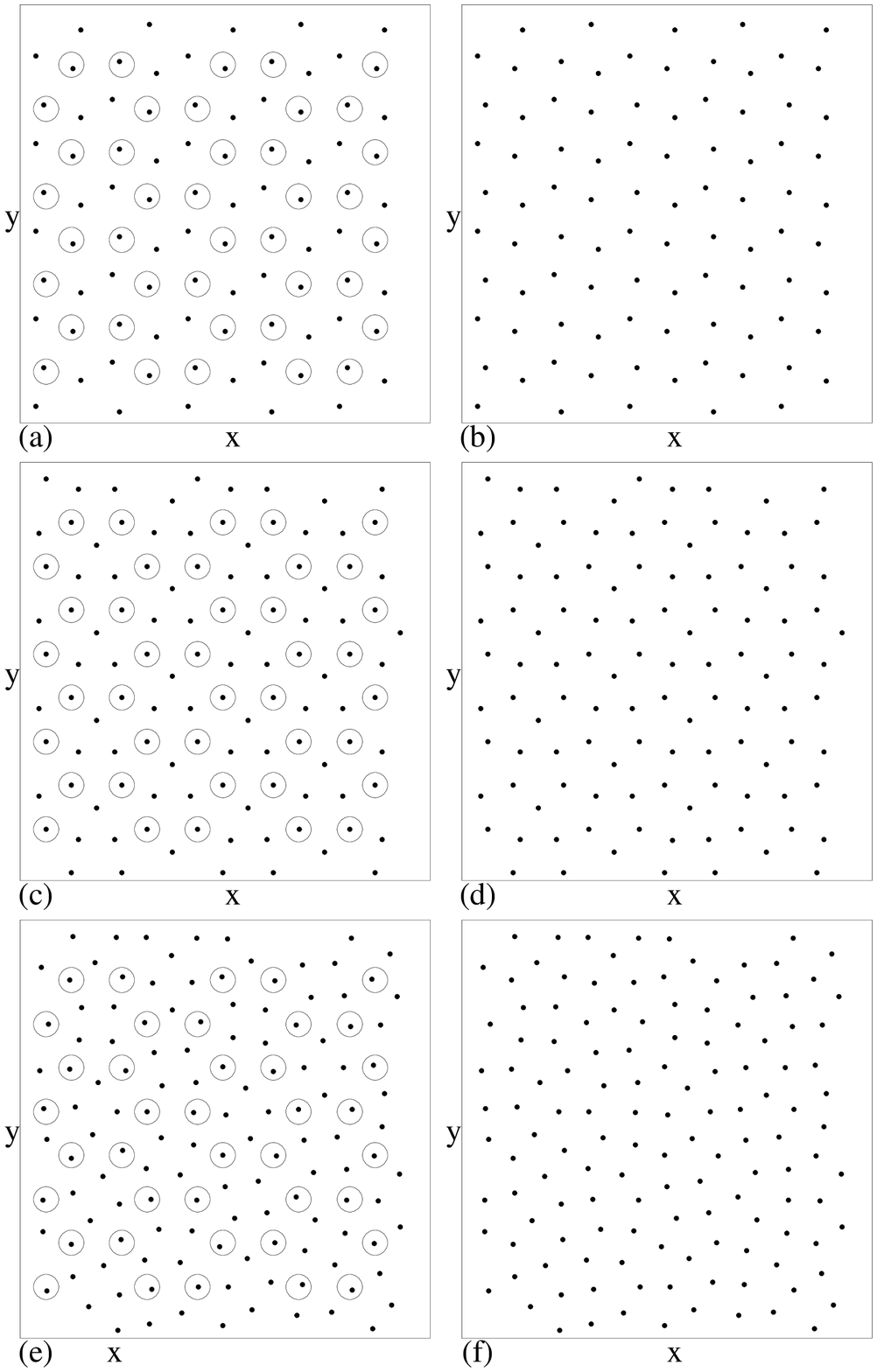}
\caption{
Left panels: Vortex positions (black dots) and pinning site positions 
(open circles) in a portion of the sample for the honeycomb 
pinning array from Fig.~\ref{fig:pinimage}(c) and Fig.~\ref{fig:depinhoney}. 
Right panels: Vortex positions only.
(a,b) $B/B^H_{\phi} = 2.0$. 
(c,d) $B/B^H_{\phi} = 2.5$. 
(e,f) $B/B^H_{\phi} = 3.0$.   
}
\label{fig:honeyimg2}
\end{figure}

\subsection{Vortex Configurations for $B> B_\phi$}  

As $B$ increases above
$B/B^H_{\phi}=1.5$, the additional
vortices sit in the large interstitial regions at the center of each
honeycomb plaquette.
In general for $1.5 < B/B^H_{\phi} < 2.0$ we 
find that the plaquette centers capture at most two vortices 
rather than three. 
The two vortices cannot both sit at the center of the interstitial site,
so instead they form an interstitial dimer state, as illustrated in 
Fig.~\ref{fig:honeyimg2}(a,b) 
at $B/B^H_{\phi} = 2.0$. 
The interstitial dimers have an additional orientational ordering which
is highlighted in Fig.~\ref{fig:honeyimg2}(a).
The dimers can be described as having a director field which
is oriented at a $30^{\circ}$ angle with the $x$ axis 
in Fig.~\ref{fig:honeyimg2}(a,b).
Within the center interstitial
region of each honeycomb plaquette,
the six surrounding pinned vortices create a sixfold 
modulated symmetric potential and each vortex forming the interstitial dimer
sits in one of the minima of this potential.
As a result, at zero temperature
each dimer can be oriented in one of three degenerate directions.  If
we change the initial conditions during the simulated annealing process
slightly, we find the same configuration shown in Fig.~\ref{fig:honeyimg2}(a,b)
with one-third probability.  With equal probabilities we obtain configurations
in which all the dimers are either oriented at $90^{\circ}$ 
or $-30^{\circ}$ angles with the $x$ axis.
If the dimers in neighboring plaquettes did not interact with each other, 
we would expect to find a random distribution of dimer orientations among
the three degenerate directions in a given configuration.
The orientational ordering  of the dimers in our system indicates that 
dimers in neighboring plaquettes do interact with each other, and that this
interaction 
gives the dimers a ferromagnetic-like alignment.
Unlike an Ising model in zero magnetic field which has
two possible spin orientations, 
this system has three possible orientations for the dimers
and is thus more closely related to a three-state Potts model. 
The ordering of the dimers is very similar to the 
ordering found in the recently studied model 
of colloidal dimers on a triangular 
lattice, which has been shown to map to the three-state Potts 
model \cite{Frey}.
The theoretical work in Ref.~\cite{Frey} indicates that 
the dimers lose their orientational ordering as a function of temperature
through either a continuum melting transition or a first order transition, 
depending on the system parameters. 
The colloidal dimers in neighboring plaquettes 
have been shown to interact through an effective quadrupole moment
with an additional screening term \cite{Trizac}; higher order $n$-mers
were also considered which interact through higher order pole moments.
These types of interactions are anisotropic and thus the pole moment 
energy can be minimized when the $n$-mers form an orientationally
ordered state.  Neighboring dimers may be oriented parallel or perpendicular
to each other depending on the pinning geometry \cite{Colloid}.
The colloids interact via a repulsive screened Yukawa potential; since
this is similar to the vortex interaction of Eq. 2,
the same type of multipole interactions 
between vortices in neighboring plaquettes should emerge as in the 
colloid system. 

For fields $ 2.0 < B/B^H_{\phi} < 2.5$, the additional vortices 
again occupy the interstitial sites where they form trimer states. 
For this range of fields we do not find any interstitial sites
that have captured four vortices.
Figure~\ref{fig:honeyimg2}(c) shows the vortex 
configuration at
$B/B^H_{\phi} = 2.5$ where each pinning site captures 
one vortex and each center interstitial region captures three vortices. 
The trimers are equilateral triangles and
each vortex in the trimer is located at
one of the sixfold potential minima created by the six 
surrounding pinned vortices.  
In the same manner as the dimer states, the trimers are orientationally
ordered and all align in one of the two possible degenerate
orientations, indicating that neighboring trimers have an interaction with
ferromagnetic character.
The overall vortex lattice structure at 
$B/B^H_\phi=2.5$ 
is very intricate, as indicated in Fig.~\ref{fig:honeyimg2}(d). 
It can be viewed as triangles of vortices each surrounded by
three pentagons, indicating that a large fraction of the system has
fivefold ordering coexisting with true long range order.

Figure~\ref{fig:honeyimg2}(e) illustrates the vortex 
configurations and pinning site locations at $B/B^H_{\phi} = 3.0$.
At this filling a weak peak in $f_{c}/f_{p}$ appears, as shown in
Fig.~\ref{fig:depinhoney}.
Here each pinning site captures one vortex and the center 
interstitial regions capture four vortices in a quadrimer state.
Unlike the dimers and trimers, 
we find no orientational ordering
of the quadrimers. 
The fourfold symmetry of the quadrimer state cannot match the sixfold
modulation of the potential at the center of the honeycomb plaquette.
If the four vortices in the quadrimer 
sit at minima of this potential, then at least two vortices
must occupy adjacent minima, which is energetically unfavorable due
to the vortex-vortex interactions.
Instead, the quadrimers form distorted square structures that are
not commensurate with the underlying sixfold modulated potential.
Since the overall orientational order of the quadrimers is absent at the
$B/B^H_\phi=3.0$ filling, 
a strong commensurate peak in $f_{c}/f_{p}$ does not occur.
A weak peak does appear at this filling,
as seen in Fig.~\ref{fig:depinhoney},
due to the fact that each interstitial region captures exactly four vortices 
at $B/B^H_\phi=3.0$.  Just above or below this field, vacancies or 
interstitials in the form of threefold or fivefold occupied interstitial
sites occur which are less strongly pinned and cause a reduction in $f_c$.

\begin{figure}
\includegraphics[width=3.5in]{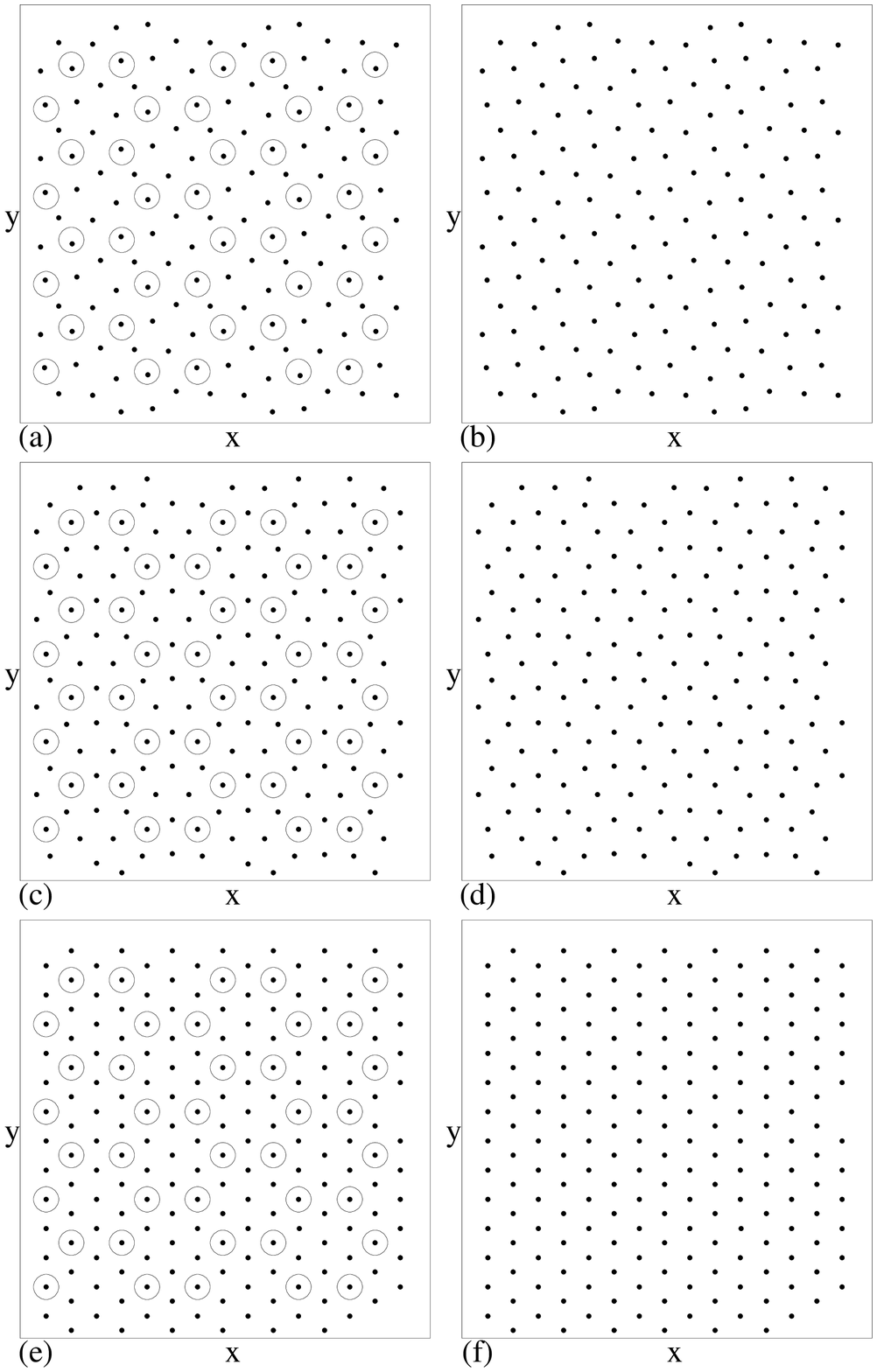}
\caption{
Left panels:
Vortex positions (black dots) and pinning site positions (open circles) 
in a portion of the sample for
the honeycomb array from Fig.~\ref{fig:pinimage}(c) and 
Fig.~\ref{fig:depinhoney}. 
Right panels: Vortex positions only.
(a,b) $B/B^H_{\phi} = 3.5$. 
(c,d) $B/B^H_{\phi} = 4.0$. 
(e,f) $B/B^H_{\phi} = 4.5$.  
}
\label{fig:honeyimg3}
\end{figure}

In Fig.~\ref{fig:honeyimg3}(a,b) we show
the vortex 
configurations at $B/B^H_{\phi} = 3.5$ 
where a strong peak in $f_{c}/f_{p}$ appears in Fig.~\ref{fig:depinhoney}(b). 
At this field, the system deviates from
the pattern 
which we observed at lower fields for $n=1$ to 4 
of forming symmetrical $n$-mers in the
interstitial sites at the center of the honeycomb plaquettes. 
Instead of forming a pentamer state at $B/B^H_\phi=3.5$, the
interstitial vortices 
are arranged with four
vortices captured as a rectangular quadrimer in the center of each interstitial plaquette
and a fifth vortex between two adjacent
pinned vortices.
The vortex lattice has long range order with all the center quadrimers aligned 
in the same direction,
unlike the $B/B^H_\phi=3.0$ filling of Fig.~\ref{fig:honeyimg2}(e,f).
The ordering is possible because
the fifth
interstitial vortex which sits at the boundary of the honeycomb plaquette
breaks the sixfold symmetry of the potential minima inside the
center of the plaquette and replaces it with a twofold symmetry which
can be matched by the remaining quadrimer of interstitial vortices.
   
At $B/B^H_{\phi} = 4.0$, a small commensuration peak 
appears in Fig.~2(b). The corresponding vortex 
configuration
is illustrated in Fig.~\ref{fig:honeyimg3}(c,d).
The vortex lattice has long range order.  Each interstitial site
captures six vortices in the form of
an 
inverted triangular structure 
with three interstitial vortices at the top, 
two in the middle and one at the bottom.    
The triangles have a twofold degenerate ordering, and we have also observed
the other possible ordering in which each triangle points upward instead
of downward along the $y$ direction.
Although the vortex lattice structure has long range order at 
$B/B^H_\phi=4.0$, 
there is not a large increase in $f_c/f_p$ at this field.
This may be due to the fact that the triangular configuration of 
interstitial vortices is unstable to fluctuation effects, as we will describe
in Section V.

In Fig.~\ref{fig:honeyimg3}(e,f) we illustrate the vortex 
configuration at $B/B^H_{\phi} = 4.5$ where
a pronounced peak in the depinning force occurs in 
Fig.~\ref{fig:depinhoney}(b). In this case 
the seven interstitial vortices captured in 
each interstitial
site are arranged with one vortex in the center 
surrounded by six interstitial vortices
sitting in the sixfold potential minima created by the six 
neighboring pinned vortices. 
The overall vortex lattice has a triangular ordering as seen in 
Fig.~\ref{fig:honeyimg3}(f), 
and the vortex configuration is the same as that which would result
at $B/B_{\phi} = 3.0$ 
for a triangular pinning array that is rotated by 90$^\circ$ relative to the
triangular lattice in Fig.~\ref{fig:pinimage}(a),
where the center interstitial vortex would be located in a pinning site. 
For $B/B^H_{\phi} = 5.0$ (not shown) we 
find a state without orientational 
order and there is no particular peak in $f_{c}/f_{p}$ at this field. 
We note that for very high applied fields $B/B^H_\phi \gg 1$, the
depinning response is dominated by shearing motion of the interstitial
vortices.

As we have demonstrated, the honeycomb array can stabilize various types 
of interstitial vortex $n$-mer states at the center of each honeycomb
plaquette.
These $n$-mers have a tendency to align 
in the same direction, indicating effective ordering
of a ferromagnetic nature which is
similar to the ordering observed for 
colloidal $n$-mer states on periodic substrates. 
At fields $B/B^H_{\phi}=n/2$ 
where an ordered vortex crystal forms 
we observe peaks in the depinning force. 
As we describe in further detail in Sec. V, 
we find that at finite and increasing temperature, 
a melting transition can occur in which the $n$-mers lose their
orientational ordering but 
remain trapped in the center interstitial regions of each honeycomb
plaquette, similar to the vortex dimer states 
in kagom{\' e} arrays 
\cite{Dominguez} and the colloidal
$n$-mer molten states \cite{Colloid,Brunner,Frey}. 
This suggests that many of the spin
model analogies developed to describe 
orientational ordering of colloidal $n$-mer states can be applied to the
vortex $n$-mers as well.
We term the orientationally ordered vortex $n$-mer states 
``vortex molecular crystals''
in analogy with molecular crystals,
which have translational order along with an additional alignment of
the molecules.
At higher temperatures where the molecules lose their
orientational ordering but remain in translationally ordered lattice 
positions, the system is referred to as a plastic crystal. 
Thus, by analogy,
states such as that in Fig.~\ref{fig:honeyimg2}(e) would be 
a vortex plastic crystal. 
The high temperature states (described in Sec. V) in which the orientational ordering has
melted at $B/B^H_{\phi} = 2$, 2.5, and $3.5$ 
would also be plastic vortex crystals.    

\section{Vortex States and Commensurability in Kagom{\' e} Arrays}

\begin{figure}
\includegraphics[width=3.5in]{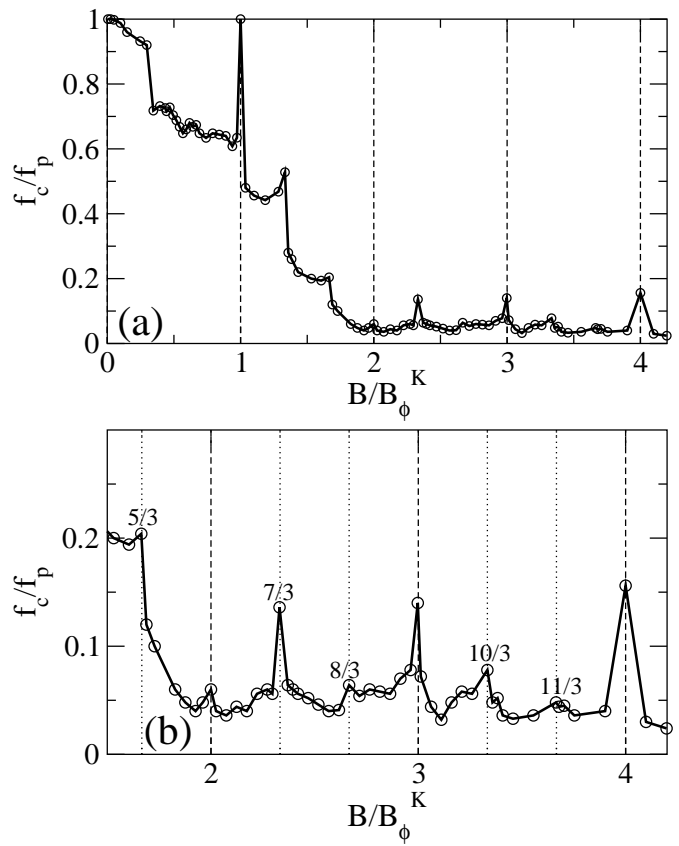}
\caption{
(a) The depinning force $f_{c}/f_{p}$ vs $B/B^K_{\phi}$ 
for the kagom{\' e} pinning array illustrated in 
Fig.~\ref{fig:pinimage}(b) with 
$f_p=0.5f_0$,
$B_\phi=0.47\phi_0/\lambda^2$, and $B^K_\phi=0.3525\phi_0/\lambda^2$. 
(b) A blowup from (a) of $f_c/f_p$ vs $B/B^K_\phi$
in the region $B/B^K_{\phi} > 2.0$.
}
\label{fig:depinkagome}
\end{figure}

\subsection{Commensurability Peaks at Integer and 
$B/B^K_\phi=n/3$ Fillings}

In Fig.~\ref{fig:depinkagome}(a) we plot $f_{c}/f_{p}$ 
obtained for vortices interacting with the kagom{\' e} 
pinning array illustrated in Fig.~\ref{fig:pinimage}(b)
with 
$f_p=0.5f_0$, $B_\phi=0.47\phi_0/\lambda^2$, and 
$B^K_\phi=0.3525\phi_0/\lambda^2$,
where
$B^K_\phi$ is the field at which the number of vortices equals the
number of pinning sites in the kagom{\' e} array. 
Unlike the honeycomb array, there is no strong peak in $f_{c}/f_{p}$ 
at $B/B^K_{\phi} = 1/2$ or at 
any of the submatching fields $B/B^K_\phi<1$ for the kagom{\' e} array, but
clear commensurability peaks occur at the integer
matching fields $B/B^K_{\phi} = 1$, 2, 3, and $4$. 
In addition to the integer
peaks there are a series of peaks 
at $B/B^K_{\phi} = n/3$ for $n > 3$. 
The strongest of these peaks fall at 
$B/B^K_\phi=4/3$, 5/3, 7/3, and 10/3, 
as shown in Fig.~\ref{fig:depinkagome}(b).
The existence of clear fractional matching effects at fields above the first
matching field is similar to the behavior seen in the honeycomb pinning
array (Fig.~\ref{fig:depinhoney}) and distinct from the behavior
of a triangular pinning array (Fig.~\ref{fig:honeytri}), where
no strong fractional matching effects appear above $B/B_\phi=1$.
This indicates that the pinning behavior of the kagom{\' e} and
honeycomb 
pinning arrays is
very similar and
suggests that similar types of 
orientationally ordered vortex molecular crystal states are occurring
in the kagom{\' e} array as in the honeycomb array. 

\begin{figure}
\includegraphics[width=3.5in]{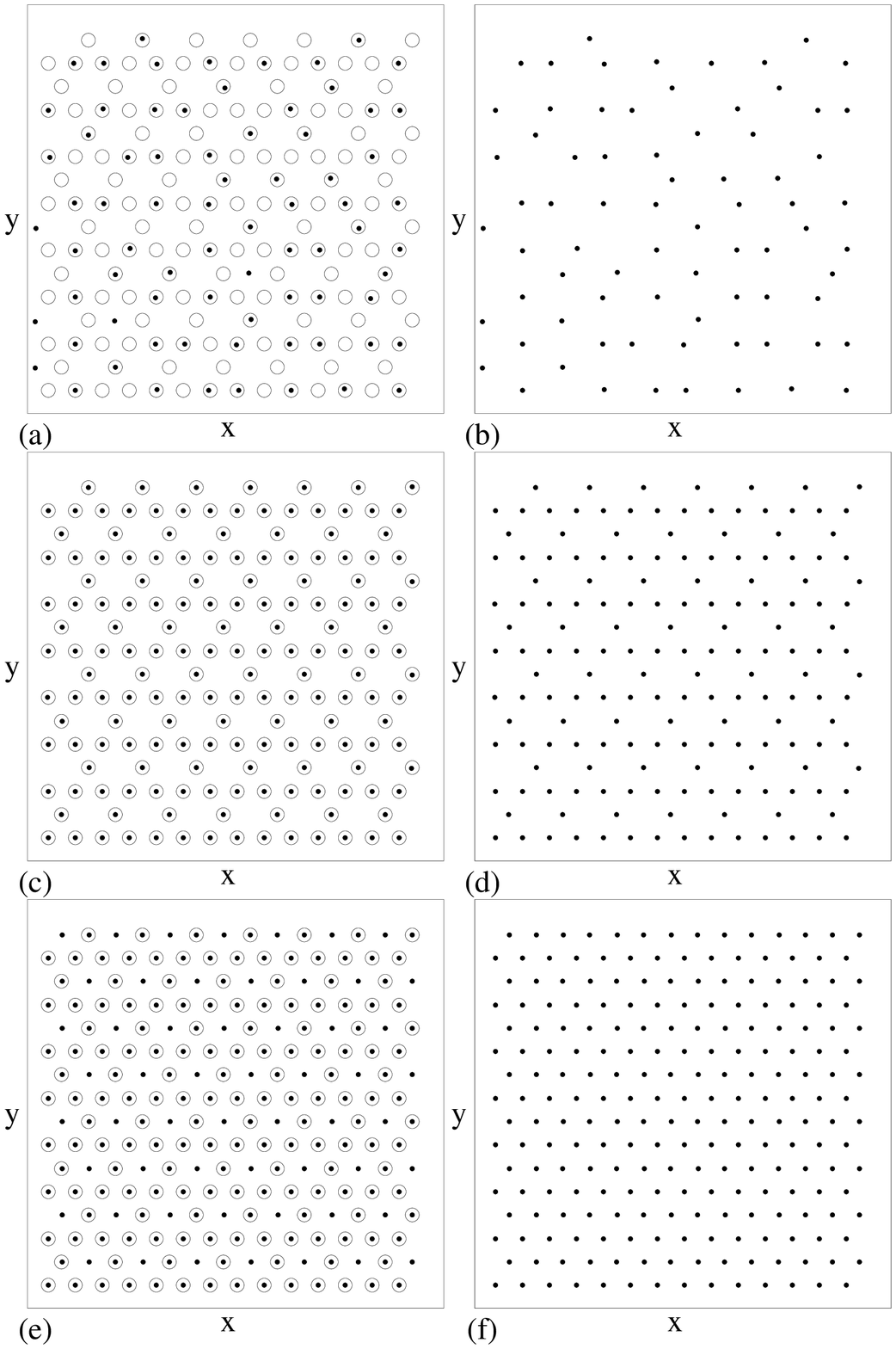}
\caption{
Left panels:
Vortex positions (black dots) and pinning site locations (open circles) 
obtained for the
kagom{\' e} array in Fig.~\ref{fig:pinimage}(b) and Fig.~\ref{fig:depinkagome}. 
Right panels: Vortex positions only.
(a,b) $B/B^K_{\phi} = 1/3$. 
(c,d) $B/B^K_{\phi} = 1.0$.
(e,f) $B/B^K_{\phi} = 4/3$. 
}
\label{fig:kagomeimg}
\end{figure}

\subsection{Vortex Configurations for $B\le B_\phi$}

In Fig.~\ref{fig:kagomeimg}(a,b) 
we show the vortex 
configuration
for the kagom{\' e} pinning array
from Fig.~\ref{fig:depinkagome} 
at $B/B^K_{\phi} = 1/3$. 
Here the  vortex lattice does not order and in general we do not observe
any particularly ordered vortex lattices for the sub-matching fields 
$B/B^K_\phi<1$ for the kagom{\' e} pinning arrays,
in contrast to the triangular and honeycomb pinning arrays. 
In the honeycomb pinning array, Fig.~\ref{fig:depinhoney}(a) showed that
there is a peak in $f_c/f_p$ at 
$B/B^H_\phi=1/3$ followed by a drop in $f_c/f_p$.
For the kagom{\' e} pinning, Fig.~\ref{fig:depinkagome} indicates that
although there is no peak in $f_c/f_p$ at 
$B/B^K_\phi=1/3$, there is still a drop in $f_c/f_p$ at this field.
In each case, the drop in $f_c/f_p$ occurs due to a change in the
nature of the vortex-vortex interactions.
For fields at and below 1/3 filling, the spacing between adjacent vortices
is at least $2a_0$ and the vortex-vortex interactions are minimal.
For fields above 1/3 filling, in order for all of
the vortices to occupy pinning sites, some of the vortices must be
located at a distance of a single lattice constant $a_0$ from another
vortex, while the spacing between other nearest-neighboring vortices
remains the larger distance $2a_0$. A vortex which has some nearest 
neighbors at a distance $a_0$ and other nearest neighbors at a distance
$2a_0$ experiences an asymmetric vortex-vortex interaction force.  This
asymmetry causes the vortex to depin at a significantly lower driving
force, producing the drop in $f_c/f_p$ above 1/3 filling.

At $B/B^K_{\phi} = 1.0$ 
where each pinning site captures one vortex, the net vortex
symmetry is that of a kagom{\' e} lattice, as indicated in 
Fig.~\ref{fig:kagomeimg}(c,d). 
At $B/B^K_\phi=4/3$, shown in
Fig.~\ref{fig:kagomeimg}(e,f), 
each center interstitial        
site of the kagom{\' e} plaquettes
captures one vortex so that the overall vortex configuration is 
triangular.
This vortex configuration is the same as that for 
the triangular pinning lattice in Fig.~\ref{fig:pinimage}(a)
at $B/B_{\phi} = 1$, where
the vortices at interstitial locations in the kagom{\' e} lattice would
sit in pinning sites in the triangular pinning lattice.
The $B/B^K_\phi=4/3$ filling also corresponds to the 
$B/B^H_\phi=1.5$ filling of the honeycomb array where each 
interstitial site captures one vortex, and to the
``first'' matching field of Ref.~\cite{Dominguez}. 
For the kagom{\' e} lattice at
$B/B^K_\phi \ll 1$, single vortex depinning dominates the response of
the system and $f_c/f_p \approx 1$. 

\subsection{Vortex Configurations for $B > B_\phi$}

\begin{figure}
\includegraphics[width=3.5in]{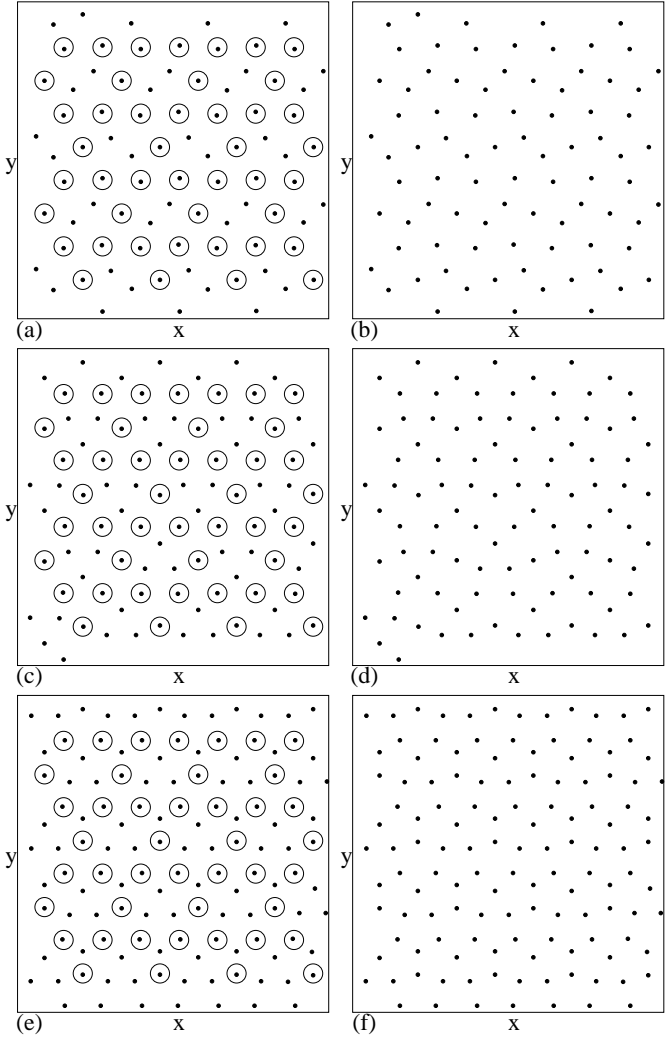}
\caption{
Left panels:
Vortex positions (black dots) and pinning site locations (open circles) 
in a portion of the sample for the
kagom{\' e} array from Fig.~\ref{fig:depinkagome}. 
Right panels: Vortex positions only.
(a,b) $B/B^K_{\phi} = 5/3$. 
(c,d) $B/B^K_{\phi} = 2.0$. 
(e,f) $B/B^K_{\phi} = 7/3$. 
}
\label{fig:kagomeimg2}
\end{figure}

Figure~\ref{fig:kagomeimg2}(a,b) illustrates the vortex configuration
at $B/B^K_{\phi} = 5/3$ in the kagom{\' e} pinning array.
At this filling, each interstitial site at the center of the
kagom{\' e} plaquettes captures two vortices 
which form a dimerized state as highlighted
in Fig.~\ref{fig:kagomeimg2}(a). 
Unlike the dimerized state in the honeycomb pinning array
where the dimers are all aligned in the same direction, 
in the kagom{\' e} array the dimers are tilted in opposite directions
from one row to the next.
This type of dimer structure is referred to as a 
herringbone state and has been observed
for colloidal dimers on triangular arrays \cite{Colloid,Frey} 
as well as for the deposition of molecular dimers on 
triangular substrates \cite{Harris}
and in three-state Potts models \cite{Vollmayr}. 

At $B/B^K_\phi=2.0$ each central interstitial site 
captures three vortices which form a trimer state, 
as shown in Fig.~\ref{fig:kagomeimg2}(c,d).
The trimer orientation is twofold degenerate 
as in the honeycomb array, and the trimers are oriented
pointing either up or down.
Fig.~\ref{fig:kagomeimg2}(c)
shows that the up-down ordering
is neither random nor uniform
but occurs
in domains reminiscent of an Ising model in zero field where 
domain walls have formed. 
If the domain wall energy is low, the relaxation time to reach
a state where one of the phases dominates can be 
exceedingly long, and here we find that it 
is beyond our computational time scale.        
In spite of the lack of true long range order,
a small peak still appears in $f_{c}/f_{p}$ at this filling,
as seen in Fig.~\ref{fig:depinkagome}(b).

In Fig.~\ref{fig:kagomeimg2}(e,f) at 
$B/B^K_{\phi} = 7/3$, 
the large interstitial sites in the kagom{\' e} plaquette centers again
capture three vortices and the additional fourth interstitial vortex 
is located in the small interstitial space between three pinning sites. 
Only half of the small interstitial regions capture a vortex. 
The trimers in  the large interstitial sites are all
aligned in the same direction, in contrast to the mixture of alignments
found at $B/B^K_{\phi} = 2.0$ in Fig.~\ref{fig:kagomeimg2}(c,d). 
The presence of the fourth interstitial vortex in the small interstitial
sites breaks the sixfold symmetry of the large interstitial region and
produces only a single low-energy alignment direction for the
interstitial trimer.
A grain boundary would require a shift in the position of the vortices
in the small interstitial sites as well as a change in the orientation
of the trimers.  Such a grain boundary has a high energy and is not
stable in our simulation.
Since there are no grain boundaries present,
this state has long range orientational order, 
and as a result the peak in $f_{c}/f_{p}$
at 
$B/B^K_{\phi} = 7/3$ 
shown in Fig.~\ref{fig:depinkagome}
is higher than the peaks at $B/B^K_{\phi} = 2.0$ and 
$B/B^K_\phi=8/3$ where 
the vortices do not form completely 
orientationally ordered states.

\begin{figure}
\includegraphics[width=3.5in]{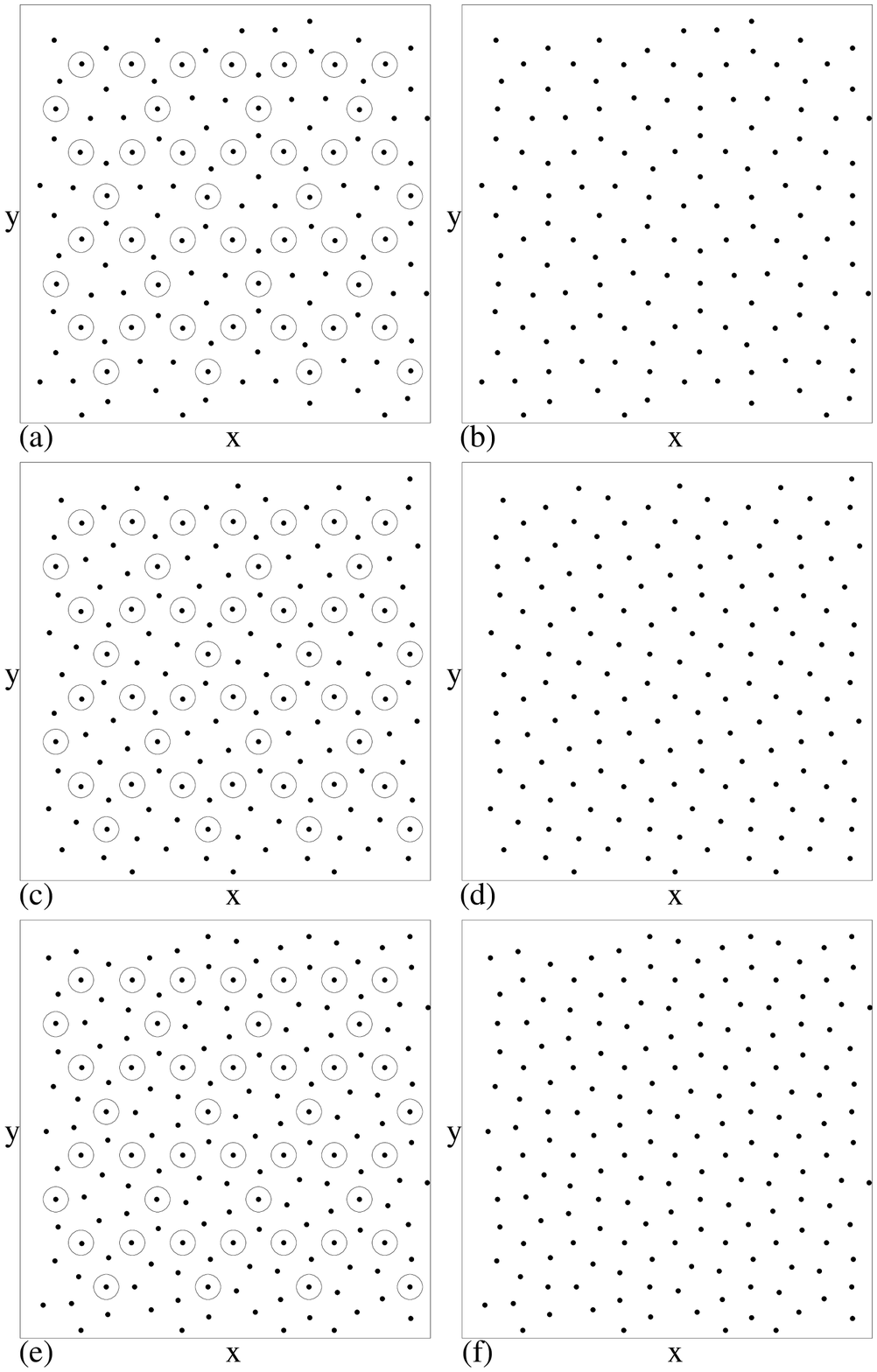}
\caption{
Left panels:
Vortex positions (black dots) and pinning site locations (open circles) 
in a portion of the sample for the kagom{\' e} array 
from Fig.~\ref{fig:depinkagome}. 
Right panels: Vortex positions only.
(a,b) $B/B^K_{\phi} = 8/3$. 
(c,d) $B/B^K_{\phi} = 3$. 
(e,f) $B/B^K_{\phi} = 10/3$. 
}
\label{fig:kagomeimg3}
\end{figure}

A related state appears at 
$B/B^K_\phi=8/3$, 
shown in 
Fig.~\ref{fig:kagomeimg3}(a,b),
where the large interstitial site
captures three vortices in a trimer state 
and all, rather than half, of the small interstitial sites capture
one vortex. 
This state was termed the ``second'' matching field in Ref.~\cite{Dominguez}.
Since all of the small interstitial sites are now occupied, the 
sixfold symmetry of the potential in the large interstitial sites is
restored, resulting in two degenerate orientations for each trimer. 
The interaction between neighboring trimers is not strong enough to
induce a global orientation of the trimers, so 
as shown in Fig.~\ref{fig:kagomeimg3}(a,b) there is no long range
trimer orientational order at 
$B/B^K_\phi=8/3$, 
just as there was no
long range orientational order at $B/B^K_\phi=2.0$ in 
Fig.~\ref{fig:kagomeimg2}(c,d).
There is also only a small peak in 
$f_{c}/f_{p}$ at 
$B/B^K_\phi=8/3$ 
as seen in Fig.~\ref{fig:depinkagome}(b).  

For $B/B^K_{\phi} = 3.0$, 
illustrated in Fig.~\ref{fig:kagomeimg3}(c,d), 
the large interstitial region captures 
four vortices which form a quadrimer state. 
The remaining interstitial vortices fill all of the 
small interstitial sites. The quadrimers are orientationally
ordered and have an effective herringbone
structure in which the quadrimers are tilted in opposite directions
from one row to the next. 
This state might more appropriately be called a double herringbone, since
the herringbone state 
is composed of single dimers,
while here two dimers
have been paired to form a quadrimer. 
As mentioned earlier, herringbone states appear in surface physics 
for the deposition of molecular dimers on 
triangular substrates \cite{Harris}. We are not aware of any observation
of the double herringbone structure for surface ordering. 
Such states might occur for the deposition of cubic molecules on 
triangular surfaces.
Fig.~\ref{fig:depinkagome} shows that there is a
prominent peak in $f_c/f_p$ at $B/B^K_{\phi} = 3.0$ 
due to the overall orientational ordering of the quadrimer state. 
    
An ordered state also forms at 
$B/B^K_{\phi} = 10/3$ as shown in 
Fig.~\ref{fig:kagomeimg3}(e,f). 
Each large interstitial region captures five vortices which form a
pentagon and the smaller interstitial sites
all capture one vortex.  
The interstitial pentagons are aligned in a single direction, 
as indicated in Fig.~\ref{fig:kagomeimg}(e). 
In surface physics it is uncommon to consider
fivefold symmetric molecules that form a pentagon; however, 
if such ring type molecules occur and lie flat on an atomic surface, then 
a similar ordering may appear.

\begin{figure}
\includegraphics[width=3.5in]{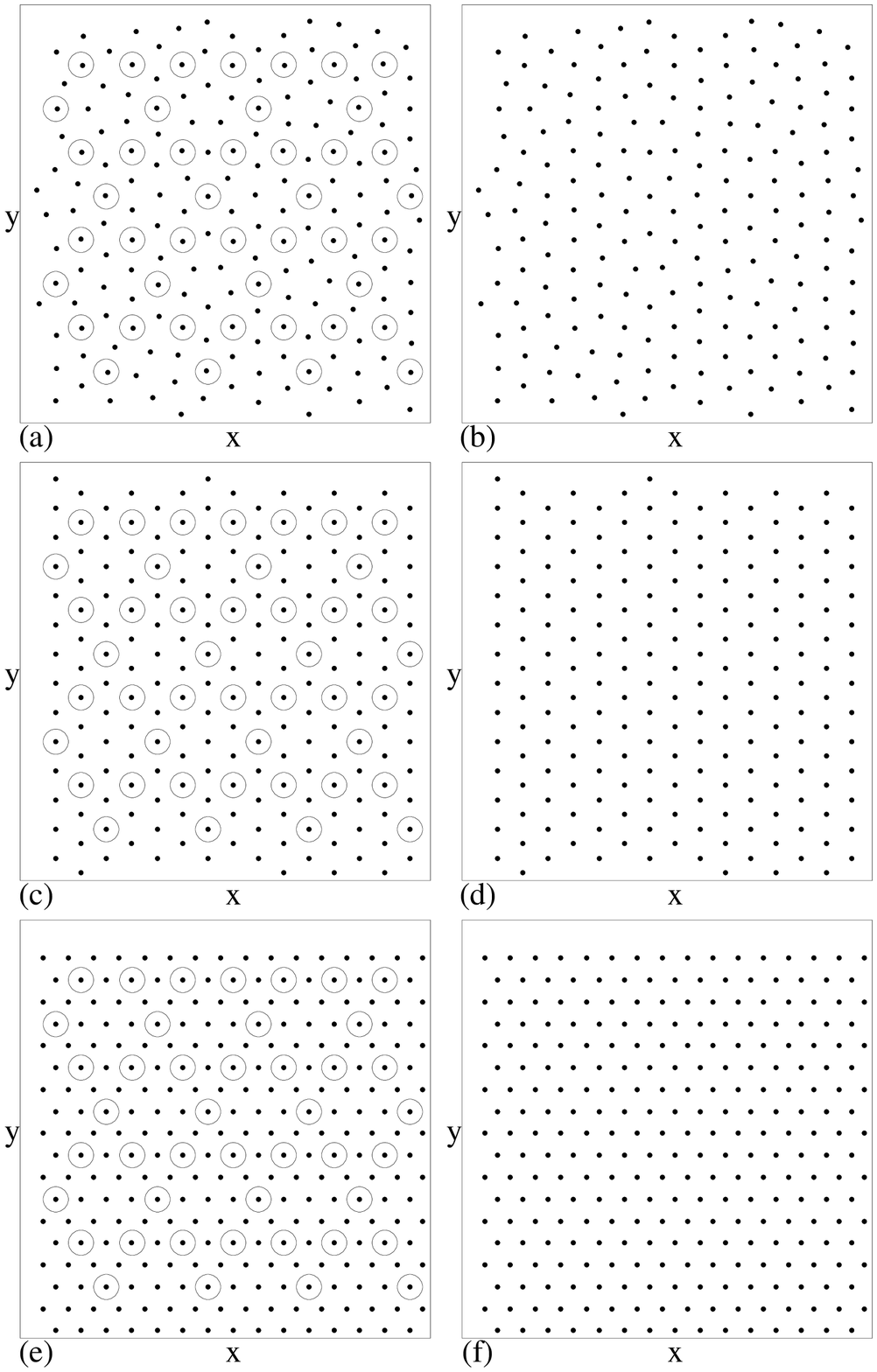}
\caption{
Left panels:
Vortex positions (black dots) and pinning site locations (open circles) 
in a portion of the sample for the kagom{\' e} array from 
Fig.~\ref{fig:depinkagome}.
Right panels: Vortex positions only.
(a,b) $B/B^K_{\phi} = 11/3$. 
(c,d) $B/B^K_{\phi} = 4$.
(e,f) $B/B^K_{\phi} = 16/3$. 
}
\label{fig:kagomeimg4}
\end{figure}

There is no appreciable peak in $f_c/f_p$ at 
$B/B^K_\phi=11/3$ in
Fig.~\ref{fig:depinkagome}.  The vortex configuration corresponding to
this field is plotted in Fig.~\ref{fig:kagomeimg4}(a,b).
All of the small interstitial sites are occupied while
some of the large interstitial sites capture seven interstitial vortices
and the rest capture five vortices.  The resulting vortex structure
has no long range orientational order.  
In general we do not observe states where six vortices are captured 
in the large interstitial regions. The triangular six-vortex configuration
that appears in the honeycomb pinning lattice in Fig.~\ref{fig:honeyimg3}(c) 
at $B/B^H_\phi=4$ is
not stable for the kagom{\' e} pinning lattice due to the presence of
the vortices in the small interstitial sites.
At $B/B^K_{\phi} = 4.0$, which was termed the ``third'' matching
field in Ref.~\cite{Dominguez}, Fig.~\ref{fig:kagomeimg4}(c,d) 
shows that each large interstitial region 
captures seven vortices and the remaining interstitial vortices fill the
small interstitial sites.
The overall vortex lattice is triangular and a prominent 
peak in $f_{c}/f_{p}$ is observed at this filling as seen in 
Fig.~\ref{fig:depinkagome}. 
For higher fillings we find additional
ordered and disordered states and in Fig.~\ref{fig:kagomeimg4}(e,f) 
we show the case of 
$B/B^K_{\phi} = 16/3$ 
where a triangular vortex lattice forms.     
At very high fillings $B/B^K_\phi \gg 1$, depinning occurs via a shearing
motion of the interstitial vortices.

These results indicate that 
the kagom{\' e} pinning array produces peaks in the critical current 
at most fields $B/B^K_\phi=n/3$ for $n > 3$. 
In some cases such as 
$B/B^K_{\phi} = 8/3$ and
$B/B^K_{\phi} = 11/3$, 
the peaks are missing or strongly reduced
due to a lack of orientational ordering in the vortex configuration. 
The general behavior for both the honeycomb and
kagom{\' e} pinning arrays is very similar 
in that $n$-mer states with various types of orientational ordering
form in the large interstitial regions of the pinning lattice.

\section{Vortex Plastic Crystal States}

\begin{figure}
\includegraphics[width=3.5in]{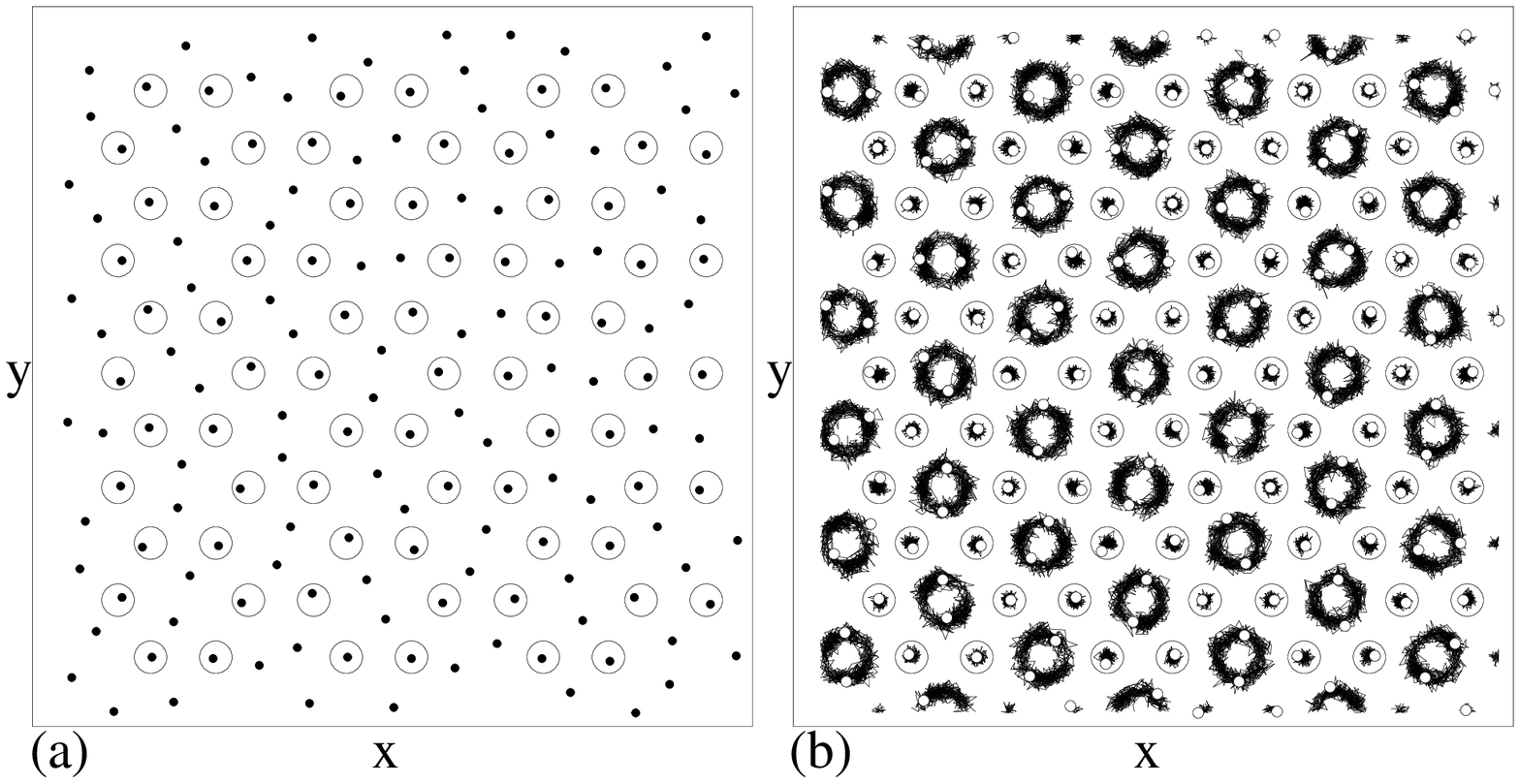}
\caption{
(a) The vortex positions (black dots) and pinning
site locations (open circles) in a portion of the sample for the 
honeycomb pinning array in Fig.~\ref{fig:honeyimg2}(a,b) with $f_p=0.5f_0$
and $B/B^H_{\phi} = 2.0$ at a temperature of $F^T = 1.56$. 
(b) The same as in (a) with vortices (white dots), pinning sites (open
circles), and vortex trajectories (black lines) over a fixed time period 
of $10^3$ simulation time steps.
}
\label{fig:meltimg}
\end{figure}

As noted in Section III(C), the
orientationally ordered dimer and higher order $n$-mer states
that form in the large interstitial regions of the honeycomb and
kagom{\' e} pinning lattices
are very similar to the colloidal molecular crystals 
studied recently \cite{Colloid,Brunner,Frey}.
Colloidal molecular crystals undergo
a thermal disordering transition to a state in which
the orientational ordering of the $n$-mers is lost when the $n$-mers
remain localized at the lattice sites but
begin to rotate freely 
\cite{Colloid,Brunner,Frey}.  
Here we show that the vortex $n$-mers exhibit 
a similar rotational melting behavior. 

In Fig.~\ref{fig:meltimg}(a) 
we plot the vortex positions and pinning site locations
for the honeycomb pinning array
in Fig.~\ref{fig:honeyimg2}(a,b)
at $B/B^H_{\phi} = 2$ and 
temperature $F^T=1.56$. 
At $F^T=0$ in Fig.~\ref{fig:honeyimg2}(a), the dimers are all aligned in a 
single direction, but at $F^T=1.56$ in Fig.~\ref{fig:meltimg}(a)
the dimer alignment is lost, although the dimers remain confined
to the interstitial sites. 
The vortex trajectories, illustrated in Fig.~\ref{fig:meltimg}(b) for
a period of $10^3$ simulation time steps, indicate that 
while the pinned vortices move a small amount, each
dimer is
undergoing rotational motion between the three low energy orientations.
We find that there is some correlation in the dimer 
motions; however, over long distances
the true long range orientational ordering is lost. 
We refer to this finite temperature state with no orientational order
as a ``vortex plastic crystal'' since it is similar
to the colloidal plastic crystal phases. 
In the vortex dimer case, there are two species of vortices: the interstitial
dimers and the single vortices trapped at the pinning sites.
In the colloidal system of Refs.~\cite{Colloid,Brunner,Frey}, the 
egg-carton substrate
does not allow interstitial colloids to exist
so there is only a single species of $n$-mer states.

\begin{figure}
\includegraphics[width=3.5in]{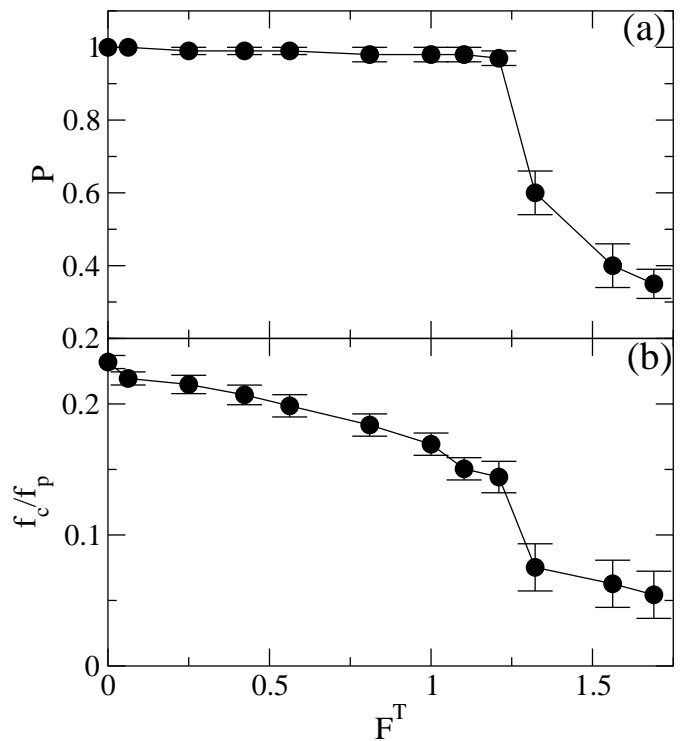}
\caption{
(a) Fraction $P$ of orientationally ordered 
dimers versus temperature $F^T$ for the honeycomb 
pinning lattice in Fig.~\ref{fig:honeyimg2}(a,b) with $f_p=0.5f_0$ and
$B/B^H_{\phi} = 2.0$. (b) The 
depinning force $f_{c}/f_{p}$ vs $F^T$ for the same sample.
}
\label{fig:orienthoney}
\end{figure}

In Fig.~\ref{fig:orienthoney}(a) 
we plot the fraction of orientationally ordered dimers 
$P$ as a function of temperature $F^T$ for 
the honeycomb pinning lattice at $B/B^H_\phi=2$ from Fig.~\ref{fig:meltimg}.
To determine $P$, for each dimer we identify the angle $\alpha$ which the line
connecting the two vortices composing the dimer makes with the 
positive $x$ axis, where $0 \le \alpha < 180^\circ$.  We then assign
a state $d_i$ to each dimer of 1, 2, or 3 depending on whether $\alpha$
is closest to $30^\circ$, $90^\circ$, or $120^\circ$.  The ordered
fraction $P$ is given by $P=N_d^{-1}\max(
\sum_{i=1}^{N_d}\delta(1-d_i),
\sum_{i=1}^{N_d}\delta(2-d_i),
\sum_{i=1}^{N_d}\delta(3-d_i))
$
where $N_d$ is the number of dimers in the sample.
For $F^T \le 1.25$ the dimers remain orientationally ordered 
with $P=1$ and form the vortex
molecular crystal state shown in Fig.~\ref{fig:honeyimg2}(a). 
For $1.25 < F^T \le 2.3$, the orientational order is thermally destroyed
and the system enters the vortex plastic crystal state. 
For $F^T > 2.3$ the dimers break apart, $P$ is no longer
defined, and there is vortex diffusion throughout the
entire system.  
 
Figure~\ref{fig:orienthoney}(b) 
shows the depinning force $f_{c}/f_{p}$ as a function of $F^T$ 
for the 
honeycomb pinning lattice at $B/B^H_\phi=2$ from
Fig.~\ref{fig:orienthoney}(a). 
The orientationally ordered
dimer state has a well defined depinning threshold which decreases 
monotonically with
increasing $F^T$.
Once the dimer orientational order is lost, 
$f_{c}/f_{p}$ undergoes a corresponding
pronounced drop. 
This is consistent with the results presented earlier in which vortex
states that lack orientational order have a significantly lower value
of $f_c/f_p$ than orientationally ordered states.
We note that in the experimental studies on honeycomb arrays 
of Ref.~\cite{Wu}, as the temperature was increased 
the matching effects at $B/B^H_{\phi} = 3.5$ and 4.5
were lost while the matching effects at 
$B/B^H_{\phi} = 1$ and 2 persisted.
This is consistent with our finding that a significant drop in
the strength of the matching effect appears at the melting of
the vortex molecular crystals, which we expect
to be present at $B/B^H_\phi=3.5$ and 4.5.

\section{Vortex Molecular Crystal Phase Diagram} 

\subsection{Honeycomb Dimer State at $B/B^H_\phi=2$}

\begin{figure}
\includegraphics[width=3.5in]{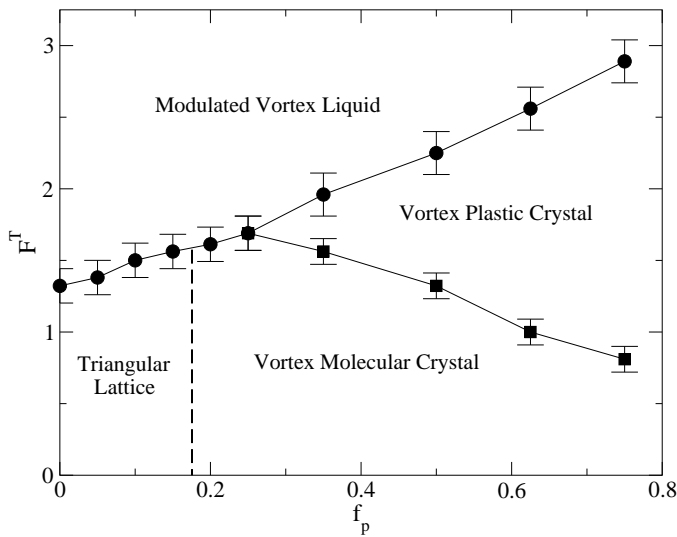}
\caption{
The temperature $F^T$ vs pinning strength $f_{p}$ phase diagram for the honeycomb pinning
array at $B/B^H_{\phi} = 2.0$ in the dimer state. 
The four phases include: the vortex molecular crystal phase
illustrated in Fig.~\ref{fig:honeyimg2}(a,b),
a vortex plastic crystal phase shown
in Fig.~\ref{fig:orienthoney}(a,b), 
a modulated vortex liquid phase where motion occurs throughout
the sample, and a partially pinned triangular vortex lattice 
that forms at low $f_{p}$ and low temperature which is described
in Fig.~\ref{fig:float}.    
Circles: Onset of significant diffusion as determined from the
diffusion measurement $D$.
Squares: Loss of orientational order as determined from 
$P$, the fraction of orientationally ordered
dimers.
}
\label{fig:phasehoney}
\end{figure}

By conducting a series of simulations and measuring $P$ and 
the diffusion $D$ we determine the phase diagram of the different
phases for the $B/B^H_{\phi} = 2.0$ dimer state of the
honeycomb pinning array. 
The diffusion is given by
\begin{equation}
D=\left\langle\frac{|{\bf R}_i(t+dt)-{\bf R}_i(t)|}{dt}\right\rangle
\end{equation}
with $dt=1000$ simulation time steps.
The resulting phase diagram as a function of temperature $F^T$
and pinning strength $f_p$ is given in Fig.~\ref{fig:phasehoney}.
The behavior of the sample at $f_p=0.5f_0$ was shown in
Figs.~\ref{fig:meltimg} and \ref{fig:orienthoney}
where a vortex molecular crystal forms for 
$0 \le F^T < 1.25$, a vortex plastic crystal is present
for $ 1.25 \le F^T < 2.3$, and at $F^T \ge 2.3$ 
the dimer states break apart and the vortices diffuse 
throughout the sample in a modulated liquid
state induced by the substrate.
Figure~\ref{fig:phasehoney} indicates that
there is no vortex plastic crystal phase for $f_p < 0.35f_0$ but that the
vortex molecular crystal melts directly to the modulated vortex liquid
for $0.15f_0 < f_p < 0.35f_0$.

\begin{figure}
\includegraphics[width=3.5in]{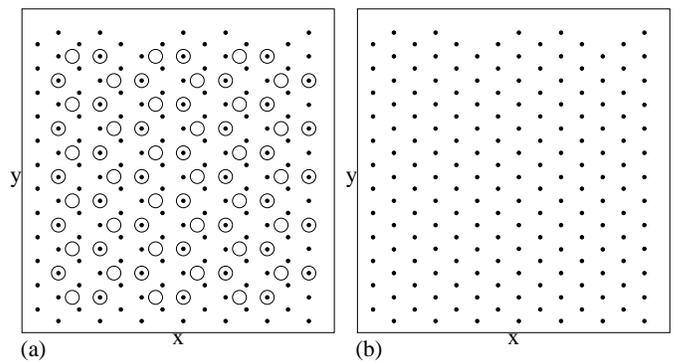}
\caption{
(a)
Vortex positions (black dots) and pinning site locations (open circles)
in a portion of the sample for the honeycomb pinning array
in Fig.~\ref{fig:phasehoney} at $B/B^H_\phi=2.0$ and 
$f_p=0.15f_0$ 
in the partially pinned triangular lattice phase at $F^T=0$.
(b) Vortex positions only.
}
\label{fig:float}
\end{figure}

For $f_p \le 0.15f_0$, 
the pinning force is not strong enough
to overcome the elastic energy of the vortex lattice and
the low temperature ground state is not the dimer
state shown in Fig.~\ref{fig:honeyimg2}(a,b) but a partially pinned
triangular vortex lattice.
This state is illustrated in Fig.~\ref{fig:float}(a,b)
at $f_{p} = 0.15f_0$ and $F^T=0$.
The triangular vortex lattice still shows a partial 
commensuration effect with the substrate and is aligned
in such a way that half of the pinning sites are occupied by a vortex.
The unpinned vortices form a kagom{\' e} structure oriented
$30^\circ$ from the original triangular lattice, where here the pinned
vortices take the place of the missing pinning sites in the kagom{\' e}
lattice.

The melting temperature 
in Fig.~\ref{fig:phasehoney} from the partially pinned triangular
lattice to the modulated vortex liquid increases with 
increasing $f_{p}$. 
The existence of pinned vortices in the triangular
vortex lattice effectively stiffens the lattice, and this effect becomes
more pronounced as $f_p$ increases.
The transition temperature 
from the vortex plastic crystal state to the modulated vortex liquid 
also increases with increasing $f_{p}$ since a higher temperature is
required to enable the vortices to escape from the stronger pinning sites.

An interesting feature in the phase diagram 
of Fig.~\ref{fig:phasehoney} is the fact that the transition temperature
between the vortex molecular crystal and vortex plastic crystal
states {\it decreases} with increasing $f_p$.
The same trend was observed for 
colloidal molecular crystals as a function of temperature versus 
substrate strength \cite{Colloid,Trizac,Frey}. 
In the colloidal case, each dimer is trapped in a substrate minima, and
as the substrate strength increases, 
the distance between the two colloids 
in each dimer decreases.  As a result, the strength of the effective
quadrupole interaction between the dimers decreases, 
lowering the transition temperature. 
For the case considered here, the vortex dimers in the honeycomb pinning
lattice do not sit in the pinning sites and are not directly affected by
the increase of $f_p$.  The dimers do, however, interact with the vortices
which are trapped by the pinning sites, and these pinned vortices become
less mobile and less able to fluctuate as $f_p$ increases.
At low $f_{p}$, the pinned vortices that are closest to the ends of each
dimer move to the outer edges of the pinning
sites in order to sit as far as possible from that dimer. 
This produces a flatter interstitial
confining potential along the length of the dimer 
in the center of the honeycomb plaquette  and permits the 
interstitial vortices that form the dimers to move further away from each
other.
For higher $f_{p}$, the pinned vortices are shifted toward the center of the
pinning sites and closer to the dimer.
This produces a stronger confining force on the interstitial dimer and brings
the vortices that form the dimer closer together, reducing the effective
quadrupole interaction between neighboring dimers and lowering the transition
temperature between the vortex molecular crystal and vortex plastic crystal
states.
This transition line is likely to saturate at very high $f_p$ when
the pinned vortices are constrained to sit at the very center of the pinning
sites and cannot adjust their positions in response to the orientation of
the neighboring dimers.
We note that it is possible 
that for very high values of $f_{p}$, a new interstitial liquid phase could
form in which the dimers break apart and the interstitial vortices hop 
from one interstitial region to another while the
vortices in the pinning sites remain immobile. 

It is beyond the scope of this work to determine the exact 
nature of the transitions in the phase diagram of Fig.~\ref{fig:phasehoney}; 
however, since the overall system appears to be very 
similar the colloidal case, we can argue from the results for the colloidal
system that the vortex molecular crystal to vortex plastic crystal 
transition is probably Ising-like. The vortex plastic crystal to 
modulated vortex liquid state 
transition is mostly likely an activated crossover.

\subsection{Honeycomb Trimer State at $B/B^H_\phi=2.5$}

\begin{figure}
\includegraphics[width=3.5in]{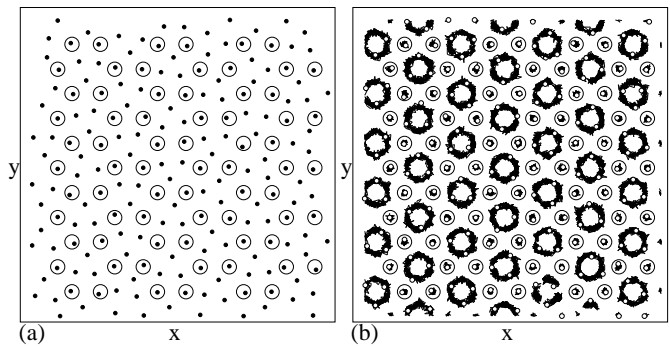}
\caption{
(a) The vortex positions (black dots) 
and pinning site locations (open circles) in a portion of the sample
for the honeycomb pinning array in Fig.~\ref{fig:honeyimg2}(c,d) with
$f_p=0.5f_0$ and $B/B^H_{\phi} = 2.5$ at a temperature of $F^T = 1.1$. 
(b) The same as in (a) with vortices (white dots), pinning sites (open
circles), and vortex trajectories (black lines) over a fixed time
period of $10^3$ simulation time steps.
}
\label{fig:melttri}
\end{figure}

In order to determine how general 
the features of the phase diagram in Fig.~\ref{fig:phasehoney} 
are, we consider the
case of the honeycomb lattice at a field of $B/B^H_\phi=2.5$
where an ordered arrangement of trimers occurs as seen in 
Fig.~\ref{fig:honeyimg2}(c,d). We find the same general melting behavior
as for the dimers. In Fig.~\ref{fig:melttri}(a) 
we illustrate the vortex plastic crystal at $F^T = 1.1$. Here each pinning site
captures one vortex and the interstitial trimers do not have 
orientational ordering. In Fig.~\ref{fig:melttri}(b) 
the vortex trajectories indicate that
the trimers are rotating in a manner similar to the dimers in 
Fig.~\ref{fig:meltimg}. For 
higher temperature $F^T$, the trimers break apart and 
we observe diffusion throughout the entire sample. 

\begin{figure}
\includegraphics[width=3.5in]{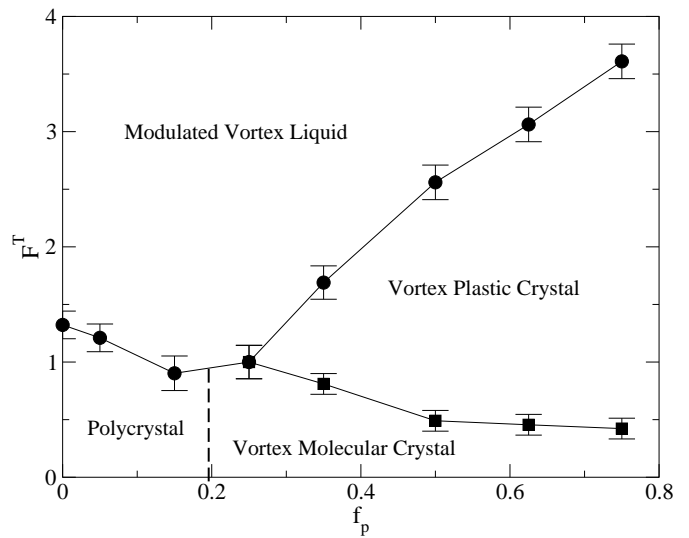}
\caption{
The temperature $F^T$ vs pinning strength $f_{p}$ 
phase diagram for the honeycomb pinning array at 
$B/B^H_{\phi} = 2.5$ in the trimer state. 
The vortex molecular crystal phase is illustrated
in Fig.~\ref{fig:honeyimg2}(c,d). 
The vortex plastic crystal phase is shown in Fig.~\ref{fig:melttri}.
In the modulated vortex liquid, there is diffusion throughout 
the entire sample,
while at low $f_{p}$ and $F^T$ a partially pinned polycrystalline 
triangular lattice forms which is described in Fig.~\ref{fig:poly}.     
Circles: Onset of significant diffusion.  Squares: Loss of trimer orientational
order.
}
\label{fig:phasetri}
\end{figure}

By conducting a series of simulations
for varied temperature $F^T$ and pinning strength $f_{p}$ and measuring the 
fraction of aligned trimers and the diffusion, we map
out the phase diagram for the trimer state 
at $B/B^H_\phi=2.5$ as shown in Fig.~\ref{fig:phasetri}.
The general features of the phase diagram at 
$B/B^H_{\phi} = 2.5$ are similar to the phase
diagram at $B/B^H_{\phi} = 2.0$ in Fig.~\ref{fig:phasehoney};
however, there
are some noticeable differences. 
For a given pinning strength, 
the trimers disorder at a 
significantly lower temperature than the dimers. 
For example, at $f_{p} = 0.5f_0$, the
trimers lose orientational order at $F^T = 0.5$ 
while for dimers the orientational order persists up to $F^T = 1.25$. 
This reflects the fact that the effective multipole interaction between trimers
is weaker than the effective quadrupole interaction between dimers.
We also note that at $F^T = 0.0$ we 
observe a peak in $f_{c}/f_{p}$ at $B/B^H_{\phi} = 2.5$, shown in 
Fig.~\ref{fig:depinhoney},
while the experiments of Ref.~\cite{Wu} did not find a peak at this
filling.
This could be due to the fact that the trimer melting temperature is
relatively low, and the experiments may have been performed above the
trimer melting temperature.

\begin{figure}
\includegraphics[width=3.5in]{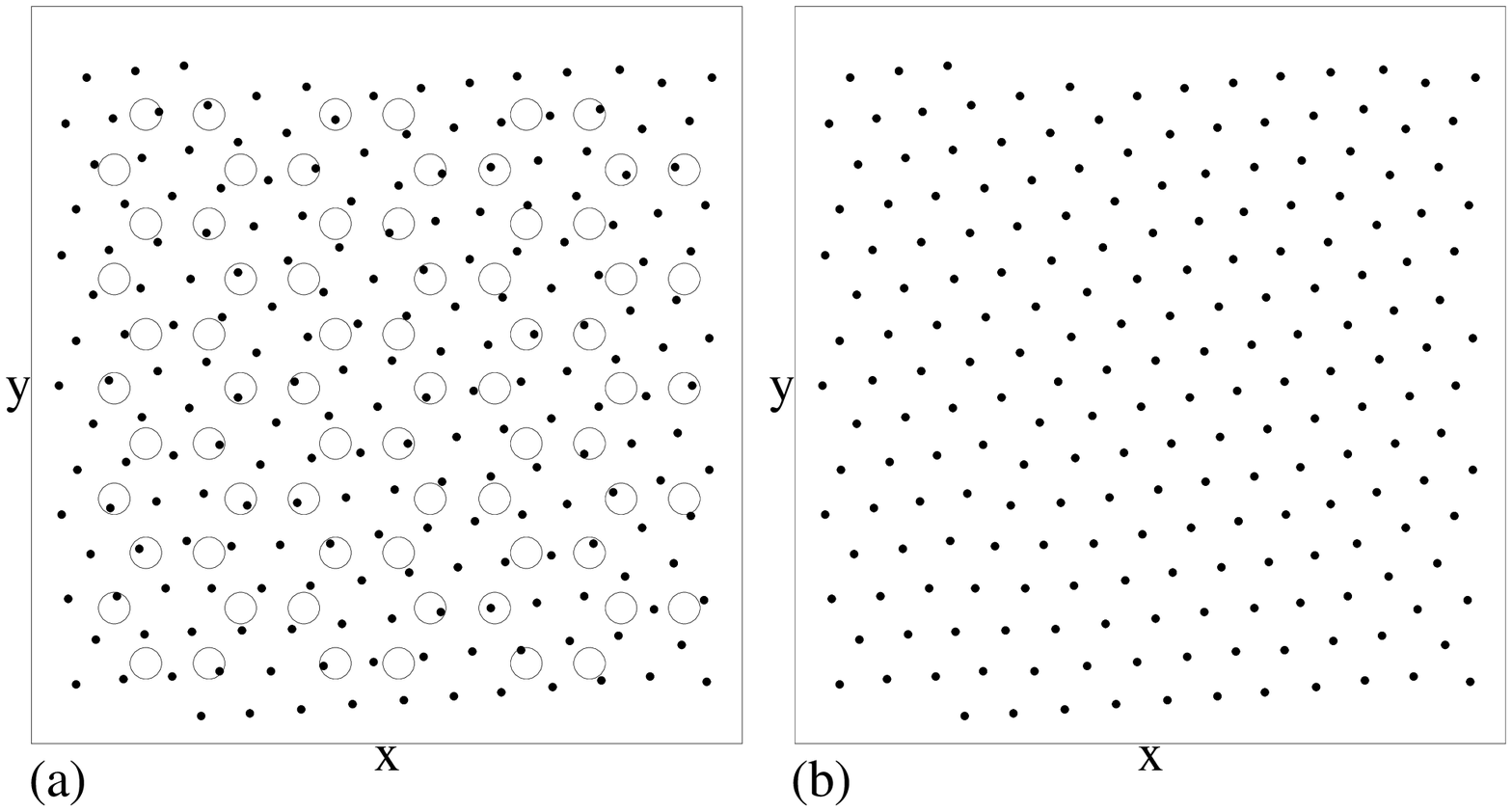}
\caption{
(a) Vortex positions (black dots) and pinning site locations (open circles)
in a portion of the sample for the honeycomb pinning array in
Fig.~\ref{fig:phasetri} at $B/B^H_\phi=2.5$ and $f_p=0.1f_0$ 
in the polycrystalline vortex state at $F^T=0$.
(b) Vortex positions only.
}
\label{fig:poly}
\end{figure}

For  $f_{p} < 0.2f_0$ and low temperatures, 
the ordered trimer state is lost and the vortices form a partially 
ordered triangular lattice as shown in Fig.~\ref{fig:poly}. 
At $B/B^H_\phi=2.5$, the vortices are unable to simultaneously
sit in a triangular lattice and align with the pinning sites, as in
the partially pinned triangular lattice of Fig.~\ref{fig:float}
at $B/B^H_\phi=2.0$.
Instead we find a polycrystalline state composed of a triangular 
lattice that contains dislocations and grain boundaries. 
The melting line between the polycrystalline state and the modulated
vortex liquid in Fig.~\ref{fig:phasetri} at $B/B^H_\phi=2.5$ decreases
in temperature with increasing $f_p$.  This is the opposite of
the behavior of the melting line between the partially pinned triangular
lattice and the modulated vortex liquid at $B/B^H_\phi=2.0$ shown in
Fig.~\ref{fig:phasehoney}, and occurs because the increasing $f_p$ leads
to an increase of the polydispersity in the lattice,
depressing the melting temperature.
At $f_{p} = 0.0$ and $B/B^H_\phi=2.5$, a dislocation-free triangular
lattice forms which has a higher melting temperature than the defected
lattice that appears at finite pinning strength.

We have also studied the effect of temperature on the orientational 
ordering of the $n$-mer states for $B/B^H_{\phi} \ge 3$ in the honeycomb
lattice at fixed $f_{p} = 0.5$ (not shown). 
In general we find that at fields with no long-range orientational order,
such as $B/B^H_\phi=3$, the $n$-mers undergo thermally induced rotations at
any finite temperature so there is only a vortex plastic crystal phase which
melts to the modulated vortex liquid.
At the other orientationally ordered fillings, we observe a finite temperature transition 
from an ordered vortex molecular crystal state to a 
vortex plastic crystal state.   

We have performed finite temperature simulations for the 
kagom{\' e} system as well and find the same general
results (not shown) as in the honeycomb system.  
At orientationally ordered fillings, there is a low 
temperature ordered vortex molecular crystal state which melts into
an intermediate vortex plastic crystal state. 
Additionally, the orientationally disordered states 
at 
$B/B^K_{\phi} = 8/3$ and $11/3$ 
show only vortex plastic crystal 
and vortex modulated liquid phases at finite temperature. 

Our results agree with
the previous study of thermal motion of 
vortices on kagom{\' e} pinning arrays.   
In Ref.~\cite{Dominguez}, 
at the second matching field $B/B_\phi=2$, 
which corresponds to our 
$B/B^K_\phi=8/3$, 
the
interstitial vortex triangles did not form an orientationally
ordered state, but instead created what 
was termed a kagom{\' e} state at finite temperatures. 
This kagom{\' e} state corresponds to our general class of 
vortex plastic crystals in which the $n$-mers
are orientationally disordered.
We note that at 
$B/B^K_{\phi} = 8/3$, 
we observe no large peak in $f_{c}/f_p$ 
as shown in Fig.~\ref{fig:depinkagome},
and the interstitial vortex trimers have no long-range orientational 
order as seen in Fig.~\ref{fig:kagomeimg3}(a). 
In contrast,
at 
$B/B^K_{\phi} = 7/3$ 
we find an orientationally ordered vortex
molecular crystal state in which all of the trimers are aligned, 
shown in Fig.~\ref{fig:kagomeimg2}(e), which has
a finite temperature transition to a vortex plastic crystal.       

The vortex plastic crystals have similarities to the 
recently proposed vortex Peierls states for vortices
in a Bose-Einstein condensate interacting with a co-rotating 
periodic optical lattice \cite{Demler}. The optical
lattice structure is in fact a kagom{\' e} 
array; however, in Ref.~\cite{Demler}
there is an additional potential minima imposed on the
center of the kagom{\' e} plaquettes.
The vortex trimer state which forms at a filling of $1/3$ on the
dual dice lattice tunnels
between the two degenerate configurations.
This state is very similar to the vortex trimer state we observe in which
the trimers are undergoing thermally induced rotations,
as shown in Fig.~\ref{fig:melttri}.         

\section{Effect of Pinning Strength}

The strength $f_p$ of the pinning sites determines 
whether it is possible for a vortex molecular crystal state to form.
In Sec.~VI we showed that in the honeycomb pinning lattice
at $B/B^H_{\phi} = 2.0$, low $f_p$ and low temperature, 
the partially pinned triangular vortex lattice 
illustrated in Fig.~\ref{fig:float}
appears 
instead of the 
vortex dimer molecular crystal shown in Fig.~\ref{fig:honeyimg2}(a).
This results when the elastic energy of the vortex lattice, which favors
a triangular vortex configuration, overcomes the energy of the pinning sites.
Similarly, at $B/B^H_{\phi} = 2.5$ 
the partially pinned polycrystalline vortex lattice seen 
in Fig.~\ref{fig:poly}
forms instead of a vortex trimer molecular crystal at low $f_{p}$. 
In general, 
partially pinned phases will not occur in superconductors with arrays of
holes because the pinning is too strong, but may form for the weaker
pinning found for vortices in superconductors with blind
hole arrays 
or for colloids on optical trap arrays.
Previous work has shown that a competition between the symmetry of the
pinning lattice and a triangular lattice of interacting particles 
can lead to structural
transitions of the particle lattice.
Transitions from square to partially pinned vortex lattices
have been observed in vortex simulations \cite{Jensen,Pogosov}, while 
partially pinned phases have recently been demonstrated experimentally
for macroscopic Wigner crystals in 
square pinning arrays \cite{Guthermann}. 
As the pinning strength is reduced at the first matching filling, 
Ref.~\cite{Guthermann} shows
that a transition occurs from a square Wigner lattice 
where each pinning site captures one charge to 
a partially pinned triangular lattice.
A similar transition appears at the second matching filling. 
If the substrate had been triangular, then at the first matching 
filling there would have been no structural transition of the charge
lattice since its symmetry would coincide with that
of the pinning lattice.
At the second matching filling on a triangular array, the particles form a
honeycomb lattice when the pinning is strong \cite{Reichhardt}, but as
the pinning strength is weakened, a transition to a partially pinned phase
should occur that allows the particle lattice to have a more 
triangular ordering. 
For vortices in honeycomb and kagom{\' e} pinning arrays,
we expect a transition from a completely
pinned phase to a partially pinned phase to occur as a function of pinning
strength even at the first matching field.  

\begin{figure}
\includegraphics[width=3.5in]{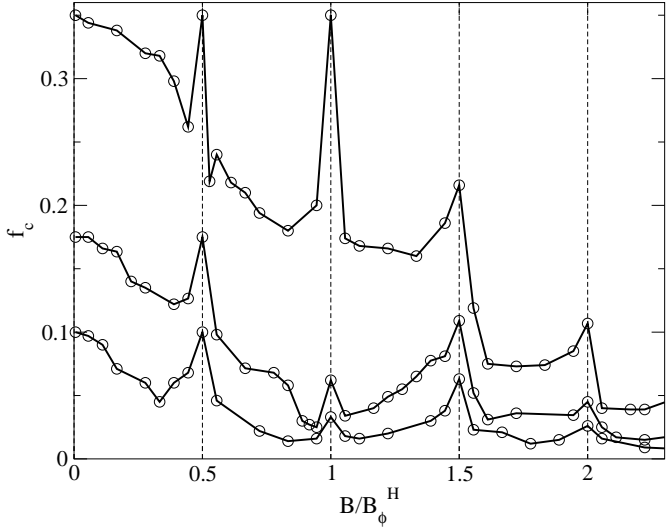}
\caption{
$f_{c}$ vs $B/B^H_{\phi}$ for the honeycomb pinning array 
from Fig.~\ref{fig:depinhoney} with
$f_{p} = 0.35f_0$, $0.175f_0$, and $0.1f_0$, from top to bottom. 
The results for $f_p=0.5f_0$ appear in Fig.~\ref{fig:depinhoney}.
}
\label{fig:pinstrength}
\end{figure}

\begin{figure}
\includegraphics[width=3.5in]{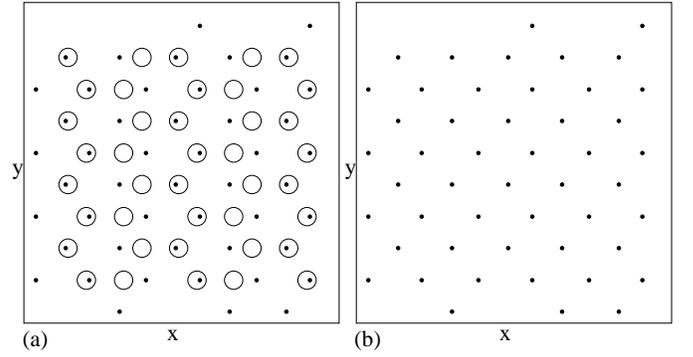}
\caption{
(a) The vortex positions (black dots) and pinning site locations (open circles) 
in a portion of the sample for the honeycomb pinning array in
Fig.~\ref{fig:pinstrength} at $B/B^H_{\phi} = 1.0$ and $f_{p} = 0.175f_0$
where a distorted square vortex lattice forms. 
(b) Vortex positions only.
For higher $f_{p}$, all the vortices are located at pinning sites 
and the state shown in Fig.~\ref{fig:honeyimg}(c,d) occurs. 
}
\label{fig:square}
\end{figure}

Since there are a rather large number of possible different vortex 
configurations shown in Sec. III and Sec. IV,
we focus here on three cases which indicate the general behavior 
for both the honeycomb and kagom{\' e} arrays at the different matching 
fields as the pinning strength is varied. 
In Fig.~\ref{fig:pinstrength} we plot $f_{c}$ vs $B/B^H_{\phi}$ 
at pinning strengths of $f_p=0.1f_0$, $0.175f_0$, and $0.35f_0$ for 
the honeycomb pinning array from Fig.~\ref{fig:depinhoney}, which
contains 
the results at $f_{p} = 0.5f_0$. 
The vortex configurations at $f_p=0.35f_0$ are the same as those at
$f_p=0.5f_0$ described in Section III, and
the depinning force $f_c$ at $B/B^H_{\phi} = 1/2$ and $1.0$ is higher 
than $f_c$ at $B/B^H_{\phi} = 1.5$ and $2.0$. 
As $f_{p}$ is reduced below $0.35f_0$, the overall $f_{c}$ at 
all the fields decreases, but not uniformly,  
as $f_{c}$ at $B/B^H_{\phi} = 1.0$ drops below the value of 
$f_c$ at $B/B^H_{\phi} = 0.5$ and $1.5$. 
This crossover in $f_{c}$ with decreasing pinning force results 
when the vortex lattice at $B/B^H_\phi=1$ undergoes a transition from a 
fully pinned honeycomb lattice such as that illustrated in 
Fig.~\ref{fig:honeyimg}(c,d) to the
partially pinned distorted square lattice shown in Fig.~\ref{fig:square}(a,b) 
at
$f_p=0.175f_0$.
Here, although there are an equal number of pins and vortices, 
half of the pinning sites are occupied while the other half are empty,
resulting in a significant decrease of the depinning threshold.
The elastic energy of the distorted square vortex lattice 
at $f_p \le 0.35f_0$ 
is 
lower than the honeycomb vortex lattice that forms
at higher $f_{p}$. 

\begin{figure}
\includegraphics[width=3.5in]{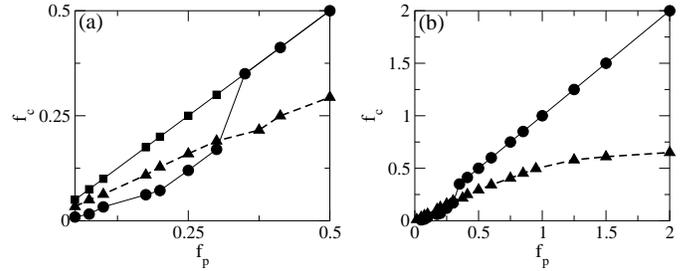}
\caption{
(a) The depinning force $f_{c}$ vs $f_{p}$ for the honeycomb pinning
array at $B/B^H_{\phi} = 1/2$ (squares), 
$B/B^H_{\phi} = 1.0$ (circles),
and $B/B^H_{\phi} = 1.5$ (triangles). 
At $f_{p} < 0.35$, the vortex lattice at $B/B^H_\phi=1$ is a distorted
square lattice rather than a honeycomb lattice, and therefore $f_c$ for
$B/B^H_\phi=1$ drops below $f_c$ for $B/B^H_\phi=1/2$.
(b) The same as in (a) 
for $B/B^H_\phi=1$ (circles) and $B/B^H_\phi=1.5$ (triangles)
but extended to higher $f_{p}$, showing that there is
a saturation in $f_{c}$ for $B/B^H_{\phi} = 1.5$.  
}
\label{fig:pinforce}
\end{figure}

In Fig.~\ref{fig:pinforce}(a) we plot $f_{c}$ vs $f_{p}$ for the 
honeycomb array at $B/B^H_{\phi} = 1/2$,
1.0, and 1.5.
At $B/B^H_{\phi} = 1/2$, the triangular vortex lattice illustrated in
Fig.~\ref{fig:honeyimg}(a,b) forms.
There is no transition to a partially pinned state as $f_p$ decreases at
this field
since there is no competition between the vortex lattice symmetry and the
pinning symmetry.  All the vortices remain trapped at pinning sites for
arbitrarily low $f_p$.
Since the vortex lattice is completely symmetrical the depinning force $f_c$
is directly proportional to $f_{p}$, 
as shown in Fig.~\ref{fig:pinforce}(a). 
At $B/B^H_{\phi} = 1.0$, all the vortices are pinned at the pinning sites in
a honeycomb lattice for $f_{p} > 0.35f_0$ 
and $f_{c}$ increases linearly with $f_{p}$. 
For $f_{p} \le 0.35f_0$ the system
enters the partially pinned phase illustrated in Fig.~\ref{fig:square}.
At the same time, $f_c$ drops abruptly to a lower value and then decreases
more slowly with decreasing $f_p$.
The vortex lattice is always triangular at $B/B^H_{\phi} = 1.5$, 
as seen in Fig.~\ref{fig:honeyimg}(e,f), and 
there is no sharp change in $f_{c}$ vs $f_{p}$
since there is no structural transition in the vortex lattice
when $f_p$ is varied. 
For $f_{p} < 0.35$, $f_{c}$  is  
higher at $B/B^H_{\phi} = 1.5$ than at 
$B/B^H_\phi=1.0$, as also indicated in Fig.~\ref{fig:pinstrength}. 
The depinning force at $B/B^H_{\phi}=1.5$ 
increases more slowly than linearly with $f_p$,
which is more clearly seen in Fig.~\ref{fig:pinforce}(b) 
where the range of $f_{p}$ is extended to $2.0f_0$. 
Here, the depinning force is determined by the 1/3 of the vortices which
are not confined at pinning sites.
As $f_{p}$ increases, the vortices at 
the pinning sites are more strongly pinned;
however, the caging potential which pins the interstitial vortices saturates. 

At $B/B^H_{\phi} = 2.0$ we have
already shown in Section VI (B) 
that a transition to the partially pinned state 
illustrated in Fig.~\ref{fig:float}
occurs
as $f_p$ decreases. 
As a result, there is a jump in $f_c$ near $f_p=0.35f_0$ (not shown) 
similar to what we find for $B/B^H_\phi=1$ in Fig.~\ref{fig:pinforce}(a).
For high $f_p$, the $f_c$ versus $f_p$ curve for $B/B^H_\phi=2$ resembles
that at $B/B^H_\phi=1.5$ shown in Fig.~\ref{fig:pinforce}(b), with a
saturation in the depinning force for the interstitial vortex dimers.
We thus expect two dominant behaviors of $f_c$ versus $f_p$ at higher
matching fields.  For those fields where the overall vortex lattice
is triangular, such as $B/B^H_{\phi} = 1.5$ and 4.5 in the honeycomb pinning
array (Fig.~\ref{fig:honeyimg}(e,f) and Fig.~\ref{fig:honeyimg3}(e,f), 
respectively)
as well as 
$B/B^K_{\phi} = 4/3$, $4$, and $16/3$ 
in the kagom{\' e} array
(Fig.~\ref{fig:kagomeimg}(e,f), Fig.~\ref{fig:kagomeimg4}(c,d), and
Fig.~\ref{fig:kagomeimg4}(e,f), respectively),
$f_c$ saturates at high values of $f_p$ but has no transition at
low values of $f_p$.
For the matching fields with nontriangular vortex configurations,
a transition to a more triangular state at low $f_{p}$ 
occurs which is accompanied by a drop in $f_c$,
and in addition there is a saturation of $f_c$ at high $f_{p}$ which always occurs
in the presence of interstitial vortices. 
It is also possible that there could be more than one 
structural transition as $f_p$ decreases. 
For example, we have shown a transition from a honeycomb vortex lattice to
a distorted square lattice (Fig.~\ref{fig:square}) in the honeycomb pinning
lattice at $B/B^H_\phi=1$.
At extremely low $f_{p}$, a second transition to a completely triangular
vortex lattice should occur.

\section{Discussion} 

In this work we only considered the case where each pinning site could
capture at most one vortex;
however, other types of vortex phases and commensurability effects 
may occur in real samples if multiple vortices are captured by the pins.
If the pinning is very strong, the
first few matching fields would correspond to the presence of 
multiquanta vortices at the pinning sites and no interstitial vortices.
The vortices would start to enter the interstitial regions 
only once the pinning site saturation field is reached.
For kagom{\' e} and honeycomb lattices of this type, 
strong commensurability effects would appear at only the integer
matching fields below the pinning saturation field, and would
cross over to matching effects at the $n/2$ or $n/3$ fields once the 
pinning saturation occurs. 
Alternatively,
individually quantized vortices may form
at low fields, 
while multiply quantized vortices may appear in the pinning sites only
at high fields where the vortex lattice constant is 
small.
In such a case, noninteger matching effects would appear at low fields 
followed by a crossover to integer matching effects at higher fields.
In Ref.~\cite{Peeters}, where continuum simulations 
of square pinning arrays were performed, it was proposed
that multiple interstitial vortices 
can merge to form a single giant interstitial vortex. 
In the case of the honeycomb and
kagom{\' e} arrays, singly quantized vortices could occupy
the pinning sites while multiply quantized vortices sit in the large 
interstitial sites.
In this scenario the overall vortex lattice would be triangular and 
commensurability peaks would be observable at all 
fields $n/3$ or $n/2$. 
If multiple quantization of the interstitial vortices 
occurs, there would be no finite temperature phase
transition from a vortex molecular crystal to a vortex plastic crystal  
and all of the matching peaks would vanish at the same rate with 
temperature.  Additionally, since the vortex lattice symmetry
would be triangular at every matching field, there would be no 
missing matching peaks due to the formation of vortex plastic crystals.  
These effects are not seen in the experiment of Ref.~\cite{Wu}, suggesting
that multiple 
quantization of vortices is not occurring. 

As stated previously, many of the orientationally 
ordered $n$-mer states are related to 
various types of Ising and three-state Potts models. 
It may be possible to create
honeycomb and kagom{\' e} pinning arrays that have an additional 
anisotropy in one direction.  This would
bias some of the degenerate directions of the $n$-mer states so that 
only certain directions of the $n$-mer ordering would occur. 
Also, highly frustrated states could be created
such as incommensurate fillings composed of
mixtures of two different $n$-mer species. 
Ordered colloidal molecular crystal alloy states 
formed by mixtures of different $n$-mers have been proposed 
in Ref.~\cite{Colson}, 
and similar states may occur for the vortex system. 
It is likely that such mixtures
would have extremely long relaxation times to reach the ordered ground state. 
These mixtures might produce glassy dynamics and have interesting 
time-dependent or history-dependent properties.  

\begin{figure}
\includegraphics[width=3.5in]{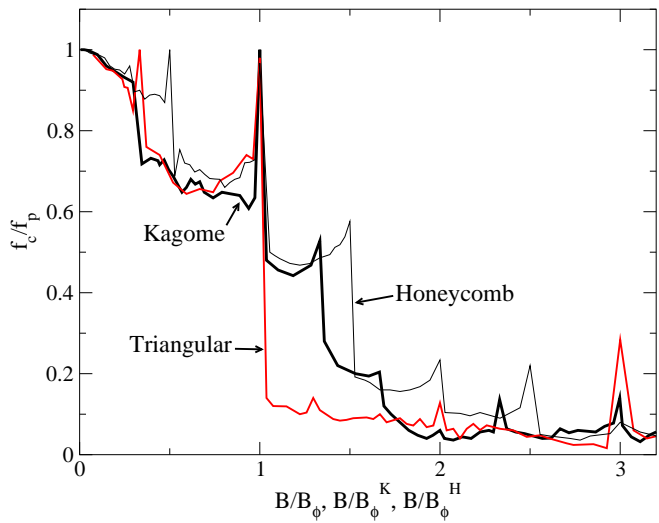}
\caption{(Color online)
$f_{c}/f_{p}$ versus $B/B_{\phi}$ for the triangular pinning array 
(red medium line),
versus $B/B^K_\phi$ for the kagom{\' e} array (dark line), and 
versus $B/B^H_\phi$ for the honeycomb array (light line) at 
$f_{p} = 0.5f_0$.
}
\label{fig:22}
\end{figure}

In addition to allowing the creation of vortex molecular crystal and
vortex plastic crystal states, 
honeycomb and kagom{\' e} pinning arrays may also be valuable for 
the general enhancement of pinning. 
If pinning sites are placed in a honeycomb
or kagom{\' e} arrangement,
higher pinning can be achieved at matching fields above the first matching
field compared to a triangular pinning array with an equal number of pinning
sites.
This is illustrated in 
Fig.~\ref{fig:22} 
where $f_{c}/f_{p}$ is plotted as a function of matching field
for a triangular, kagom{\' e}, and honeycomb pinning array
of equal pinning strength $f_p=0.5f_0$. 
Below the first matching field there is little difference between 
the depinning force for the three arrays; however, above the
first matching field the honeycomb array shows a pronounced enhancement 
of $f_c$ over the triangular array and the
kagom{\' e} array also shows a smaller enhancement
of $f_c$ compared to the triangular array. 
This suggests that for different arrays with the
same number of pinning sites, the honeycomb and 
kagom{\' e} pinning arrays have an overall higher critical
current than triangular pinning arrays.

\section{Conclusion} 

We have used numerical simulations to 
study the vortex states in honeycomb and kagom{\' e} 
pinning arrays in the limit where only one vortex is captured at 
each pinning site.
For the honeycomb arrays, we find pronounced matching effects 
in the form of peaks in the depinning force at 
most fields $B/B^H_\phi=n/2$ where $n > 2$, 
while for kagom{\' e} arrays pronounced matching effects occur 
at most fields $B/B^K_\phi=n/3$ where $n > 3$. 
This is in contrast to the purely
triangular pinning arrays which have prominent matching effects
only at $B/B_\phi=n$ or at fields below the first matching field.
For the honeycomb pinning array, a variety of novel vortex 
molecular crystal states occur.  Here, 
multiple interstitial vortices occupy the large interstitial regions
of the honeycomb pinning lattice and form effective dimer, trimer 
and higher order $n$-mer vortex states. The $n$-mers
interact via an effective quadrupole or higher order pole moment 
which can lead to an overall $n$-mer
orientational ordering.
This type of $n$-mer ordering is similar
to the recently studied colloidal molecular crystal states 
observed for repulsively interacting
colloidal particles on periodic substrates. 
For the honeycomb array, at some of the matching fields
the $n$-mers do not order and there is no prominent peak in the 
depinning force at these fields.
Our results agree well with recent experiments on honeycomb pinning
arrays where strong peaks are observed at $B/B^H_\phi=n/2$ fillings 
with missing or weak peaks for the fields at which we observe
disordered $n$-mer configurations.
For the kagom{\' e} pinning arrays we observe similar dimer, 
trimer and $n$-mer ordered states 
which form herringbone or other ferromagnetic-like configurations.
Most of the missing commensuration peaks in the 
$B/B^K_\phi=n/3$ sequence correspond to the lack of $n$-mer ordering
at zero temperature.
We term the orientationally ordered vortex $n$-mer states 
``vortex molecular crystals.''
At finite temperature we demonstrate the existence of a transition 
from an orientationally ordered
vortex molecular crystal state to a state where the 
$n$-mers are rotating and lose their relative orientational ordering. 
We refer to this disordered state as a vortex plastic crystal. 
At higher temperatures there is a crossover to a modulated
vortex liquid state where there is diffusion throughout the entire system.  
When the pinning strength is weak, we find
that the vortex molecular crystal phases 
undergo a transition to partially pinned phases 
where the vortex lattice has triangular or partially triangular 
ordering and only a portion of the pinning sites are occupied.   
These results suggest that the 
vortex molecular crystal states  have many similarities
to the recently studied colloidal molecular crystal states 
and that the vortex molecular crystal to vortex plastic
crystal transitions can be mapped to various types 
of spin systems such as Ising and Potts models.   

\section{Acknowledgments}
We thank M. Hastings for useful discussions. 
This work was carried out under the auspices of the National Nuclear
Security Administration of the U.S. Department of Energy at Los
Alamos National Laboratory under Contract No.
DE-AC52-06NA25396.

\end{document}